\documentclass[a4paper,10pt,twoside]{cpc-hepnp}
\usepackage[colorlinks,linkcolor=blue,anchorcolor=blue,citecolor=blue,urlcolor=blue,breaklinks=true]{hyperref}

\usepackage{multicol}
\usepackage{graphicx}
\usepackage{booktabs}
\usepackage{amssymb,bm,mathrsfs,bbm,amscd}
\usepackage[tbtags]{amsmath}
\usepackage{lastpage}
\usepackage{color}
\usepackage{CJKutf8}

\usepackage[T1]{fontenc} 
\usepackage{lmodern,bbm}
\usepackage{booktabs}  
\usepackage{graphicx}
\usepackage{xspace}
\usepackage[small]{subfigure}
\usepackage{color}
\usepackage{multirow}
\usepackage{marginnote}
\usepackage{slashed}
\usepackage{mathrsfs}
\usepackage{threeparttable}
\usepackage{hyperref}
\usepackage{float}

\newcommand{\nn}{\nonumber}

\newcommand{\be}{\begin{equation}}
\newcommand{\ee}{\end{equation}}
\newcommand{\bes}{\begin{subequations} \begin{align} }
\newcommand{\ees}{\end{subequations}\end{align} }
\newcommand{\bea}{\begin{eqnarray}}
\newcommand{\eea}{\end{eqnarray}}

\newcommand{\Lqcd}{\Lambda_{\text{QCD}}}

\newcommand{\as}{\alpha_s}

\newcommand{\cusp}{\mathrm{cusp}}

\newcommand{\Ga}{\Gamma}

\newcommand{\ta}{\tau_a}
\newcommand{\tmean}[2]{\langle \tau_{#1}^\text{#2} \rangle}

\newcommand{\cO}{\mathcal{O}}

\newcommand{\cL}{\mathcal{L}}

\newcommand{\eq}[1]{Eq.~\eqref{eq:#1}}
\newcommand{\eqs}[2]{Eqs.~\eqref{eq:#1} and \eqref{eq:#2}}

\renewcommand{\sec}[1]{Sec.~\ref{sec:#1}}
\newcommand{\ssec}[1]{Sec.~\ref{ssec:#1}}
\newcommand{\sssec}[1]{Sec.~\ref{sssec:#1}}
\newcommand{\appx}[1]{App.~\ref{app:#1}}
\newcommand{\fig}[1]{Fig.~\ref{fig:#1}}

\newcommand{\tab}[1]{Table~\ref{tab:#1}}

\newcommand{\cjt}{c_{\tilde{J}}^{2}}
\newcommand{\cjone}{c_{\tilde{J}}^{1}}
\newcommand{\muns}{\mu_{\mathrm{ns}} }

\newcommand{\NNLLp}{$\text{NNLL}'$ }
\newcommand{\NLLp}{$\text{NLL}'$ }
\newcommand{\omgll}{\mathcal{O}(\Lambda_\text{QCD}^2/m_H^2)}

\allowdisplaybreaks[1]

\begin{document}

\fancyfoot[C]{\small \thepage}


\begin{CJK*}{UTF8}{gbsn}
\title{Angularity in Higgs boson decays via $\boldsymbol{H\to gg}$  \\ at \NNLLp accuracy\footnote{The work of DK, JZ, and YS is supported by the National Key Research and Development
Program of China under Contracts No.~2020YFA0406301 and by the National Natural Science Foundation of China (NSFC) through Grant No.~12150610461.
The work of JG was sponsored by the National Natural Science Foundation of China
under the Grant No.~12275173 and No.~11835005.
The work of TM is supported by the Science and Engineering Research Board (SERB) 
through the SRG (Start-up Research Grant) of File No.~SRG/2023/001093.}}

\author{{Jiawei Zhu (朱佳伟)}$^1$, {Yujin Song (宋昱锦)}$^1$, {Jun Gao (高俊)}$^2$, {Daekyoung Kang}$^{1,3}$\email{dkang@fudan.edu.cn}, {Tanmay Maji}$^4$
}

\maketitle

\address{$^1$ Key Laboratory of Nuclear Physics and Ion-beam Application (MOE) and Institute of Modern Physics, \\ Fudan University, Shanghai 200433, China\\
$^2$School of Physics and Astronomy, Shanghai Key Laboratory for Particle Physics and Cosmology, and Key Laboratory for Particle Astrophysics and Cosmology (MOE), Shanghai Jiao Tong University,
Shanghai 200240, China\\
$^3$Department of Physics, Korea University, Seoul 02841, Korea\\
$^4$Department of Physics, National Institute of Technology, Kurukshetra, Haryana, 136 119, India
}

\end{CJK*}

\begin{abstract}
We present improved predictions of a class of event-shape distributions called angularity for a contribution from an effective operator $H\to gg$ in Higgs hadronic decay that suffers from large perturbative uncertainties. In the frame of Soft-Collinear Effective Theory, logarithmic terms of the distribution are resummed at \NNLLp accuracy, for which 2-loop constant of gluon-jet function for angularity is independently determined by a fit to fixed-order distribution at NLO corresponding to $\cO(\as^2)$ relative to the Born rate. Our determination shows reasonable agreement with the value in a thesis recently released. In the fit, we use an asymptotic form with a fractional power conjectured from recoil corrections at one-loop order and it improves the accuracy of determination in positive values of angularity parameter $a$. The resummed distribution is matched to the NLO fixed-order results to make our predictions valid at all angularity values.  We also discuss the first moment and subtracted moment of angularity as a function of $a$ that allow to extract information on leading and subleading nonperturbative corrections associated with gluons.
\end{abstract}



\begin{multicols}{2}

\section{Introduction}\label{sec:intro}

Since the experimental discovery of Higgs boson, the final piece of the Standard Model, the continuous operation of the Large Hadron Collider (LHC), the ATLAS and CMS experiments \cite{Aad:2012tfa, Chatrchyan:2012xdj} have shown great success on refined study of the Higgs boson, for example, determination of the Higgs couplings with top quarks \cite{Sirunyan:2018hoz, Aaboud:2018urx} and bottom quarks \cite{Aaboud:2018zhk, Sirunyan:2018kst}. 
At the LHC, the huge background restricts the accuracy of measurements on the Higgs signal strength not to below 5$\%$ \cite{CMS:2018qgz} and also very difficult to probe Yukawa couplings of the light fermions of first two generations  \cite{Gao:2013nga, Soreq:2016rae, Bishara:2016jga, Bodwin:2013gca, Kagan:2014ila, Zhou:2015wra, Koenig:2015pha, Perez:2015lra, Chisholm:2016fzg}. The sensitivity to Higgs self-interactions is also weak \cite{Goertz:2013kp, Sirunyan:2018two, CMS:2018ccd, Aaboud:2018ftw}.
There are future experimental proposals to probe its properties and the rare decay modes of Higgs boson with higher accuracy, e.g., the International Linear Collider \cite{Behnke:2013xla} and the Circular Electron-Positron Collider (CEPC) \cite{CEPCStudyGroup:2018ghi}, CLIC \cite{Lebrun:2012hj}, FCC-$ee$ \cite{Gomez-Ceballos:2013zzn}. In the future electron-positrons colliders, most of the possible decay channels of Higgs boson can be studied with high precision and total width of the Higgs boson can be reconstructed in a model independent way. 

Hadronic decay of Higgs boson, where the final state consists of hadrons, is one of dominant decay modes that will be extensively studied in the Higgs factories.
The hadronic decay is initiated by quarks and gluons. The former is induced by Yukawa-coupling operator $H\to q\bar q$ and the latter is by an effective gluon operator $H\to gg$ through the top-quark loop, which we refer to as the quark channel and gluon channel, respectively. 
Event shapes are the classic observables designed to describe the geometry of final hadrons. 
An application of event shapes for Higgs decay is discussed in \cite{Gao:2016jcm}, where analysis with a number of event shapes enables to constrain Yukawa coupling of light quarks.
Studies of event shapes in Higgs decays includes fixed-order results at next-to-leading order (NLO) corresponding to $\cO(\as^2)$ relative to the Born rate \cite{Gao:2019mlt,Luo:2019nig,Gao:2020vyx,Coloretti:2022jcl,Gehrmann-DeRidder:2023uld}, resummed result of thrust at next-to-next-to-leading logarithmic accuracy (\NNLLp) for gluon channel \cite{Mo:2017gzp}, \NNLLp matched to the NLO \cite{Alioli:2020fzf} and approximate N$^3$LL$^{'}$ accuracy \cite{Ju:2023dfa}.
Angularity distributions are available at \NNLLp for the quark channel, at \NLLp for the gluon channel \cite{byan:cepc2022} and at NNLL in a groomed version for the gluon channel \cite{vanBeekveld:2023lsa}.

The angularity is a class of event shapes defined by a continuous weight parameter $a$ that controls sensitivity to the rapidity parameter as
\be
\ta = \frac{1}{m_H} \sum_{i} |{\bf p}^i_\perp| e^{-\left|\eta_i \right|(1-a)} 
\,,\label{eq:ta}
\ee
where $m_H$ is the Higgs mass, the sum goes over all the final particle $i$ having transverse momentum ${\bf p}^i_\perp$, and rapidity $\eta_i$ is measured with respect to thrust axis, which is an axis minimizes the value of thrust $\tau_{a=0}$ .
For the change in $a$, the angularity $\ta$ is sensitive to a collinear particle having large rapidity $|\eta|\gg 1$ and less sensitive to a soft particle having smaller rapidity $|\eta|\sim \cO(1)$. In other words, by regulating the value of $a$ one can control relative contribution between the collinear and soft particles or modes.
With its continuous parameter $a$, the angularity enables to interpolate parameter region between thrust  $(a=0)$ and jet broadening $(a=1)$ and further extrapolate beyond this region. The infrared-safe region of angularity parameter $a$ is $- \infty < a < 2$, where perturbation theory is valid. For the region $a \geq 2$, the angularity is sensitive to collinear splitting and becomes infrared-unsafe.
In the framework of soft-collinear effective theory (SCET) \cite{Bauer:2000ew,Bauer:2000yr,Bauer:2001ct,Bauer:2001yt,Bauer:2002nz}, the factorization of angularity distribution in small $\ta$ region and resummation of Sudakov logarithms of $\ta$ are established and studied in thrust-like region $a<1$  in $e^+e^-$ \cite{Bauer:2008dt,Hornig:2009vb,Bell:2018gce} and in DIS  \cite{Zhu:2021xjn}, and the broadening-like region $a \sim 1$ in $e^+e^-$ \cite{Budhraja:2019mcz,Becher:2011pf,Becher:2012qc}.  
There are also studies of variants with recoil-free axis \cite{Larkoski:2014uqa}, with vector transverse momentum \cite{Bijl:2023dux}, and with joint two angularities \cite{Procura:2018zpn}.
In this work, we focus on improved prediction of the region $a<1$.

The gluon channel starting at $\as^2$ is known to suffer from large perturbative uncertainties compared to the quark channel. 
Angularity distribution for the gluon channel is only available at \NLLp, which is one-order lower than that of quark channel \cite{byan:cepc2022}. Improving the gluon channel, resummed up to \NNLLp accuracy and matched up to next-to-leading order (NLO) accuracy, is the primary focus of this work. The most crucial part of it is to determine a remaining term, 2-loop constant of gluon-jet function in factorization and to compute the fixed-order distribution at NLO for angularity in Higgs hadronic decay. 

Our paper is organized as follows.
We first review factorization formula for angularity distribution in \sec{form}, which is used for resummation of logarithmic terms.
In \ssec{strategy} we review the determination strategy of jet-function constant that makes use of a fit to fixed-order distribution,   in \ssec{lo} we discuss the one-loop constant to illustrate a fit form with fractional power, then in \ssec{nlo} we determine two-loop constant in  the quark- and gluon-jet function using Tikhonov regularization to handle ill-posed fit.
In \sec{angdis} we present our prediction of angularity distribution resummed at \NNLLp accuracy and matched to NLO fixed-order distribution, and discuss leading and subleading nonperturbative corrections. 
Finally, we conclude in \sec{con}.

\section{Factorization and resummation}
\label{sec:form}

At the small angularity limit, the final state hadrons can be treated as nearly back-to-back jets in the rest frame of the Higgs boson.  Similar to the $e^+ e^- \to q\bar{q}$, in the framework of soft-collinear-effective theory (SCET) \cite{Bauer:2000ew, Bauer:2000yr, Bauer:2001yt, Beneke:2002ph, Beneke:2002ni}, the factorization formula for the Higgs boson decay is given by  \cite{Hornig:2009vb,Bauer:2008dt,Gao:2019mlt}
\begin{equation}
\begin{aligned}
\frac{d\Gamma^i}{d\ta }&= 
\Gamma_B^i(\mu) |C_t^i(m_t,\mu)|^2 \, |C_S^i(m_H,\mu) |^2
 \int d\ta^{J1} \, d\ta^{J2}  d\ta^S \,\\
& \times\delta \bigg( \ta - \ta^{J1} - \ta^{J2} -\ta^S \bigg)
 J^i (\ta^{J1},\mu) \, J^i(\ta^{J2},\mu) \, S^i(\ta^S,\mu) \, ,\label{eq:fac1}
 \end{aligned}
\end{equation}
where $i=q, g$ represent the quark and gluon channel, respectively. The Born decay rates are $\Gamma_B^g(\mu)=  \frac{m_H^3 \alpha^2_s(\mu) }{72 \pi^3 v^2} $ and $\Gamma_B^q(\mu)=  \frac{m_H  C_A y^2_q(\mu) }{16 \pi }  $.
The Wilson coefficient $C^g_t(m_t, \mu)$ corresponds to the top-quark loop coupled to two gluons \cite{inami1983effective,djouadi1992higgs,Chetyrkin:1997iv,Chetyrkin:1997un,Chetyrkin:2005ia,baikov2017five}, while for quark $C^q_t(m_t, \mu)=1$. 
$C^i_S(m_H, \mu)$ represents the hard coefficient that can be found by integrating out the hard fluctuations near to the hard scale $m_H$. This hard coefficients are defined from the matching to SCET and can be obtained from $Hq\bar{q}$ and $Hgg$ form factors that are available up to the 3-loop order \cite{Schwartz:2007ib, Becher:2008cf, Bauer:2008dt,Gehrmann:2010ue}. 
The angularity soft functions $S^i(\ta,\mu)$ describing soft emissions is available up to 2-loop order \cite{Bell:2018gce,Bell:2020yzz,Bell:2018oqa} for both quark and gluon, which are related by the Casimir scaling.
The angularity jet functions $J^i (\ta,\mu)$ describe the collinear emission along the direction of initial quark or gluon. The quark-jet function is known up to 2-loop, and the gluon-jet function at the 1-loop \cite{Hornig:2009vb,Bell:2018gce} is known a while ago while a remaining constant term in 2-loop gluon-jet function became available recently \cite{Brune:2022cgr}. 
Here, one of the main tasks of this paper is to independently compute the 2-loop constant by using method described in \sec{const}.

In \eq{fac1}, the convergence in perturbation theory may break down due to the presence of Sudakov logarithms of $\ta$. At small $\ta$ limit the logarithm shows singular behavior. A standard way to cure this behavior is through resummation of these large logs in momentum space \cite{Ligeti:2008ac,Abbate:2010xh} or in Laplace space \cite{Becher:2006mr,Becher:2006nr}. Here we work in the Laplace space, where the transformation and factorization in \eq{fac1} read as 
\end{multicols}
\begin{equation}
    \begin{aligned}
\tilde \Gamma^i(\nu_a)&= \int^\infty_0 d\ta e^{-\nu_a \ta }  \frac{d\Gamma^i}{d\ta }  \\
&=   \Gamma_B^i(\mu) \, |C_t^i(m_t,\mu)|^2 \, |C_S^i(m_H,\mu) |^2 
 \tilde{J}^i \left( \nu_a,\mu\right) \, \tilde{J}^i\left(\nu_a,\mu\right) \, \tilde{S}^i\left(\nu_a,\mu\right) 
 \, , \label{eq:fac2}
    \end{aligned}
\end{equation}
\begin{multicols}{2}
\noindent where $\nu_a$ is the conjugate variable, and $\tilde{J}^i$ and $\tilde{S}^i$ are the jet and soft functions in the Laplace space.

A generic form of $G=\{C_t,C_s,\tilde{J},\tilde{S}\}$ in the fixed-order expansion in the strong coupling constant $\as$ is given in \eq{str}, that are composed of cusp $\Ga(\as)$, non-cusp $\gamma(\as)$ anomalous dimensions, and constant terms. 
 The large logs present in each function $G$ are resummed by the renormalization group (RG) evolution starting from natural scales $\mu_G$ to the desired scale $\mu$, where the scales $\mu_G$ are chosen as a physical momentum that minimizes the logs. 
Details of resummation are summarized in \appx{sscet} and \appx{res}.

At each order of $\as$, log $L$ terms behave like $\as^n L^{k}$, where $0\le k\le 2n$.
In logarithmic accuracy, the power counting  of log $L$ is $\as L\sim \cO(1)$. Then, terms scaling like $\as^n L^{n+1}\sim \cO(1/\as)$ are called leading log (LL), terms like $\as^n L^{n}\sim \cO(1)$ are next-to-leading log (NLL), and terms like $\as^n L^{n-k}\sim \cO(\as^k)$ are N$^k$LL. 
At LL accuracy, the cusp anomalous dimension $\Gamma(\as)$ of each function and QCD beta function $\beta(\as)$ are needed, and at NLL accuracy, non-cusp anomalous dimensions $\gamma(\as)$ begins to contribute, and at NNLL accuracy, 1-loop constant terms are needed. At our target accuracy \NNLLp 2-loop, constant terms should be added to NNLL accuracy. \tab{logaccuracy} summarizes the logarithmic accuracy and relevant $n$-order ingredients in $\as$ such as $\Gamma(\as), \beta(\as), \gamma(\as)$ and constant terms that are given in \appx{sscet}  and \appx{anom}. 

Since our main focus is the gluon channel, computing the 2-loop constant term for the gluon-jet function and \NNLLp resummation of the gluon channel, we set $i=g$ everywhere in \eqs{fac1}{fac2} and omit the superscript from now on.

\end{multicols}
\begin{table}[htb]
 $$
\begin{array}{|c|c|c|c|c|}
\hline
 & \Gamma(\as) & \gamma(\as) & \beta(\as) & \text{constant in } \{C_t,C_s,J, S\}[\as] \\ \hline
 \text{LL} & \as & 1 & \as & 1 \\ \hline
  \text{NLL$(')$} & \as^2 & \as & \as^2 & 1 (\as)\\ \hline
    \text{NNLL$(')$} & \as^3 & \as^2 & \as^3 & \as (\as^2) \\ \hline
\end{array}
 $$
\caption{Resummation accuracy N$^k$LL and primed accuracy N$^k$LL$'$. Individual ingredients necessary at the corresponding accuracy: cusp and non-cusp anomalous dimensions, beta function, and hard, jet, and soft functions
\label{tab:logaccuracy}}
\end{table}
\begin{multicols}{2}

\section{Constant  term of angularity jet function} \label{sec:const}

In this section, we discuss a determination of the 2-loop constant term in the gluon-jet function necessary at \NNLLp accuracy.

The constant is determined using NLO fixed-order result by following the strategy in \cite{Hoang:2008fs,Becher:2008cf,Bell:2018gce}.
In the determination, we introduce a fitting function, which improves numerical accuracy of the constant. 
The fitting function takes the asymptotic form for nonsingular part used in thrust \cite{Abbate:2010xh} when $a\le 0$ and contains additional terms with fractional powers associated with recoil corrections obtained in \cite{Budhraja:2019mcz} when $a>0$.

We first make a quick review of the strategy in \ssec{strategy}. Then, we start with the one-loop constant by using LO result to illustrate and test recoil corrections in the fitting to nonsingular part in \ssec{lo}. In \ssec{nlo}, we determine the two-loop constants of the quark- and gluon-jet functions by using NLO results in the quark and gluon channels, respectively.

\subsection{Review of determination strategy}\label{ssec:strategy}

The SCET factorization formula in \eq{fac1} reproduces the singular terms of fixed-order QCD result when it is expanded and truncated at a fixed order in $\as$. The singular part can be expressed as
\begin{equation}
\frac{1}{\Ga_B} \frac{d \Gamma_{s}}{d \tau_{a}}=A \delta\left(\tau_{a}\right)+\left[B\left(\tau_{a}\right)\right]_{+}
\,,
\label{eq:singdiff}
\end{equation}
where the rate is normalized by the Born rate $\Ga_B=\Ga_B(m_H)$ at $\mu=m_H$, a constant term $A$ contains contributions from constant terms of jet function as well as soft and hard functions, and the subscript $+$ on a function $B$ implies the plus distribution in $\ta$. The term $A$ is also contained in the fixed-order QCD result, which can be expressed as sum of singular part in \eq{singdiff} and a nonsingular part $r(\ta)$ as
\be
\frac{1}{\Ga_B} \frac{d \Gamma}{d \tau_{a}}=A \delta\left(\tau_{a}\right)+\left[B\left(\tau_{a}\right)\right]_{+}+ r\left(\tau_{a}\right)
\,.\label{eq:fqcddiff}
\ee
Because the same $A$ appears in \eqs{singdiff}{fqcddiff}, the constant of jet function can be determined from the fixed-order result.
However, in practice the analytical results at NLO that can capture the delta function term \eq{fqcddiff} is not available for the angularity distribution.

We numerically compute the angularity distribution by using the codes used in predictions of thrust distribution \cite{Gao:2019mlt}, where matrix elements are generated by {\rmfamily\scshape OpenLoops}\cite{Buccioni:2019sur,vanHameren:2010cp,Ossola:2007ax} in part and the phase space is integrated by Monte Carlo (MC) simulation
with Vegas algorithm \cite{hahn2005cuba}. 
The numerical computation cannot capture term $A$ due to the delta function $\ta=0$ in \eq{fqcddiff}, so we use total decay rate $\Ga_t$ , which is the integration of \eq{fqcddiff} and given in \cite{Gorishnii:1990zu} for the quark channel and \cite{Chetyrkin:1997iv} for the gluon channel. The rate takes the following form
\begin{equation}\label{eq:tot}
\frac{\Ga_t}{\Ga_B}=\int^{1}_{0}d\ta^\prime\, \frac{1}{\Ga_B}\frac{d\Gamma}{d\ta^\prime}=A+r_c
\,,
\end{equation}
where the plus distribution $B$ in \eq{fqcddiff} vanished up on integration and $r_c$ is a reminder function $r_c(\ta,1)$ at $\ta=0$, 
that is defined by accumulated nonsingular part $r(\ta)$ from $\ta$ to 1 as
\be
r_c(\tau_a,1)=\int_{\tau_{a}}^{1} d \tau_{a}^{\prime} r\left(\tau_{a}^{\prime}\right)
=\int_{\tau_{a}}^{1} d \tau_{a}^{\prime} \frac{1}{\Ga_B} \left( \frac{d \Gamma}{d \tau_{a}^{\prime}}-  \frac{d \Gamma_{s}}{d \tau_{a}^{\prime}}\right)
\,.\label{eq:remainder}
\ee
In the last equality, $r(\tau_a)$ is estimated by taking a difference between \eqs{singdiff}{fqcddiff}.

Note that in numerical approach, we cannot compute \eq{remainder}  at $\ta=0$.
The remainder function is computed down to small $\ta \ll 1$ regions instead, and an approximate value of $r_c$ is obtained by taking a fit using an asymptotic form in the region and extrapolating the fit to $\ta=0$.   
Then, the value of $A$ can be obtained by subtracting the approximate value from \eq{tot}.
The fit method improves the accuracy of $A$ determination. For example, in thrust limit, the error of $A$ at NLO in gluon channel reduces from 2.6 \% to 0.7 \% using the fit with data above $\ta=10^{-4}$ and this corresponds to a reduction from 20 \% to 5 \% in the constant term.

\subsection{One-loop constant}
\label{ssec:lo}

In this subsection, we discuss the determination of one-loop constant by using LO QCD result in order to illustrate the fit to nonsingular part and to test the effect of recoil corrections.

\begin{figure}[tbh]
\centering
\includegraphics[scale=0.36]{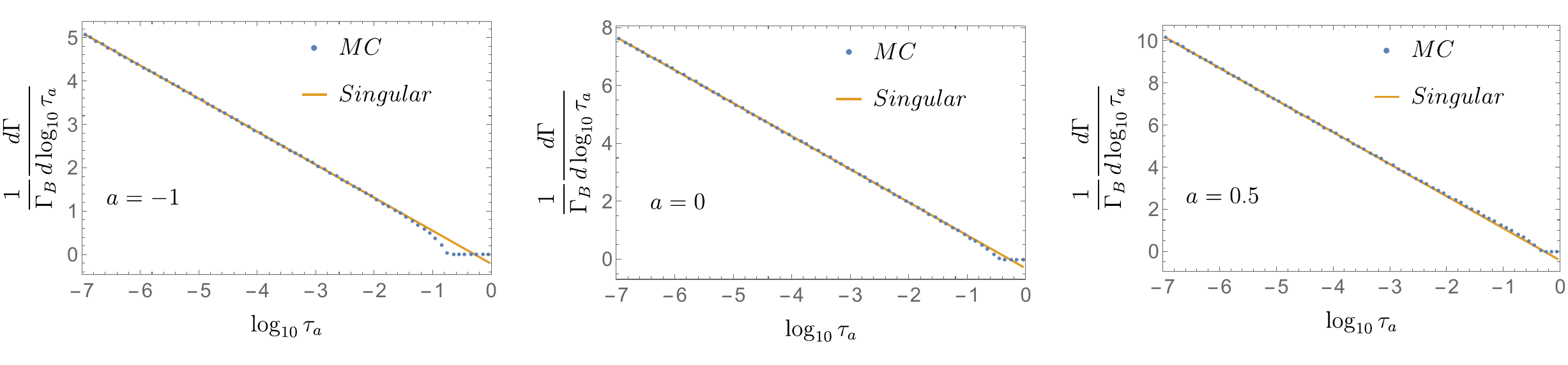}
\caption{LO MC distribution (dots) and singular distribution (solid line) in gluon channel.}
\label{fig:rate_LO}
\end{figure}

We obtain the LO angularity distribution from $10^{10}$ MC events
and bin the distribution in logarithmic space.
In \fig{rate_LO}, the MC distribution in the gluon channel is compared with singular distribution at three values of angularity parameter $a=-1, 0,$ and $0.5$. In the limit of $\tau_a\rightarrow 0$, as expected, the MC agrees well with the singular part, while in relatively large $\ta$ region they show significant differences due to large nonsingular part.

\begin{figure}[htb]
\centering
\includegraphics[scale=0.36]{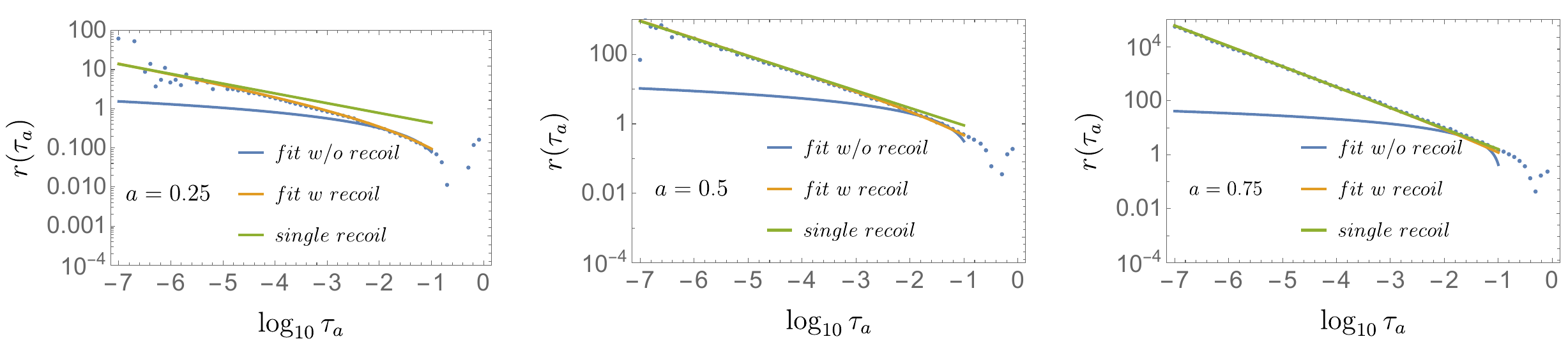}
\caption{LO nonsingular part (dots) and fit results with and without recoil corrections (blue and orange, respectively) and fit only with recoil corrections}
\label{fig:r_LO}
\end{figure}

Then, we obtain nonsingular results $r(\ta)$, as shown in \fig{r_LO}, by taking the difference between MC and singular results.
In $a\le 0$, the nonsingular $r(\ta)$ follows an asymptotic form as
\begin{equation}
\sum_{k=0}^1 \alpha_k \ln^k{\tau_a}
\label{eq:fitAsym}
  \,.\end{equation}
\end{multicols}

\begin{multicols}{2}

One can fit with above form to the nonsingular part to improve accuracy in the small $\ta$ region when the data suffers from large uncertainties.

On the other hand, for $a>0$, the nonsingular part has subleading singular terms that are more significant contributions than those in \eq{fitAsym} and become increasingly large with increasing $a$.
According to SCET power counting of angularity, collinear and soft momenta respect following scaling:
\begin{equation}
p_c^\mu \sim Q\left(\lambda^2, 1, \lambda\right) \quad \text{and} \quad p_s^\mu \sim Q\lambda^{2-a}
\,,
\label{eq:scale}
\end{equation}
where $\lambda $ is a small parameter in SCET and defined by the observable $\lambda^{2-a}\sim \tau_a$. 
The collinear momentum $p_c^\mu$ along a unit vector $\hat n$ is expressed in the light-cone coordinates and the three components are $n\cdot p_c $, $\bar n \cdot p_c$, and $p_c^\perp$ where $n=(1,\hat n)$ and $\bar n =(1, -\hat n)$.
For angularity $a\lesssim 0$, the thrust axis is insensitive to transverse recoil by soft radiation, i.e., $p^\perp_c \gg p^\perp_s$, but for $a\rightarrow 1$, the recoil effect is significant, i.e., $p^\perp_c\sim p^\perp_s$.  The recoil corrections obtained from one-loop soft function in \cite{Budhraja:2019mcz} are given by 
\begin{equation}
-\frac{\as C_i}{\pi}\frac{4}{2-a}\sum_{n=1}^{\lceil 1 / (1-a))\rceil-1} \frac{c_n}{\ta^{1-n (1-a)}}
\,,
\qquad \left( 0<a<1 \right)
\label{eq:fitRec}
\end{equation}
where the color factor $C_i=C_F$ and $C_A$ for quark and gluon channels, respectively and $\lceil x \rceil$ is the $ceiling\ function$.
The series sum turns into a single term $\tfrac{c_1}{\tau^{a}}$ for $a\in (0,1/2]$, two terms $\tfrac{c_1}{\tau^{a}}+\tfrac{c_2}{\tau^{2a-1}}$ for $a\in (1/2,2/3]$, three terms $\tfrac{c_1}{\tau^{a}}+\tfrac{c_2}{\tau^{2a-1}}+\tfrac{c_3}{\tau^{3a-2}}$ for $a\in (2/3,3/4]$, and so on.
The recoil corrections are characterized by the fractional power, which is still suppressed compared to the singular terms $\ln^n\ta/\ta$, but they are greater than typical power corrections $\ln^n\ta$ in \eq{fitAsym}. The coefficients $c_n$ for $n=1,2,3$ are given by

\begin{equation}\label{eq:coerecol}
c_1=-1, \quad c_2=\frac{1}{2}(3-2a), \quad c_3=-\frac{1}{6}\left(20-27a+9a^2\right)
\,.
\end{equation}
In our fit, the coefficients are set to free parameters that are determined by the fit to data. 
The effect of recoil corrections is shown in \fig{r_LO}, where three fit results using a form in \eq{fitAsym} without any recoil corrections (blue), a form with both \eq{fitAsym} and \eq{fitRec} (orange), and a form with just recoil terms in \eq{fitRec} without \eq{fitAsym} (green).
We find that the fitting results in orange and in green agree with \eq{coerecol}.

 At $a=0.25$, the fit with both terms in \eqs{fitAsym}{fitRec} works, while at $a = 0.5$ and $0.75$, the fit with just recoil terms in \eq{fitRec} can mostly describe the data. As $a$ increases, the recoil corrections become larger and dominate over typical power corrections in \eq{fitAsym}, and the fit without the recoil cannot describe the behavior of nonsingular data. 
\end{multicols}

\begin{figure}[!htb]
\centering
\includegraphics[scale=0.36]{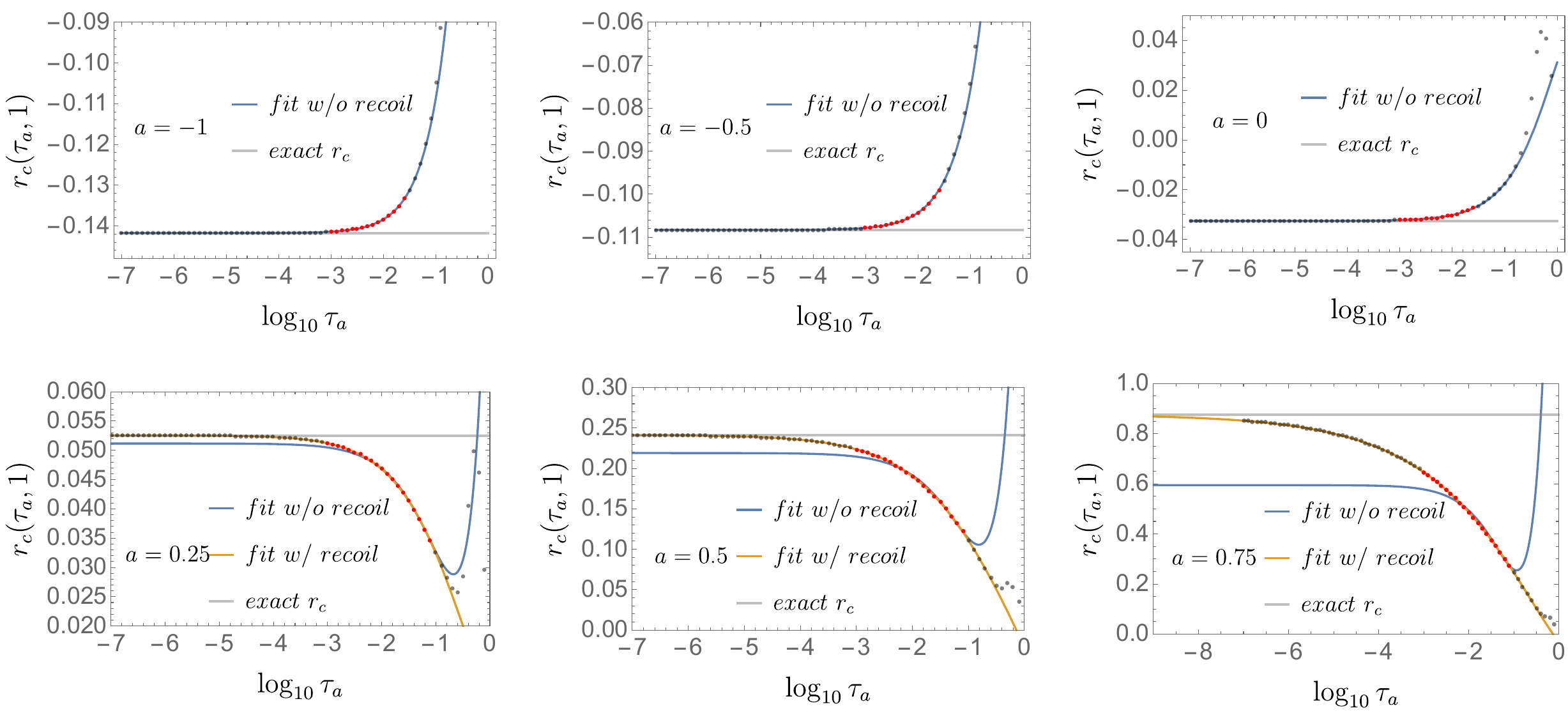}
\caption{Remainder function (dots) at LO in gluon channel, fit result with and without recoil corrections (blue and orange respectively) to data in fit regions (red).
}
\label{fig:rc_LO}
\end{figure}

\begin{multicols}{2}

\fig{rc_LO} shows the remainder function $r_c(\ta,1)$ (dots) obtained by integrating $r(\ta)$ in \fig{r_LO}, fit curves similar to \fig{r_LO}, and exact value of $r_c$ (horizontal gray line) obtained from exact singular part and LO total rate. 
In selection of fit regions (red), plateau regions are excluded in order to test the prediction of the fit result.

The fits with recoil corrections extrapolate well to the exact values of $r_c$ in $a>0$, while fits without recoil corrections extrapolate well to the exact value in $a \le 0$ and begin to deviate from the value for $a>0$.
The fact that  data for $r_c(\ta,1)$ shows plateau region, which is well aligned with the exact value of $r_c$, does not motivate such fit and extrapolation since fine-tuning errors in the difference between LO fixed-order and singular part is not significant compared to those at NLO. We again emphasize that the purpose of this subsection is to illustrate the significance of including recoil corrections shown in the fits and motivate NLO fit with those terms.  

\end{multicols}
\begin{figure}[!tbh]
\centering
\includegraphics[scale=0.33]{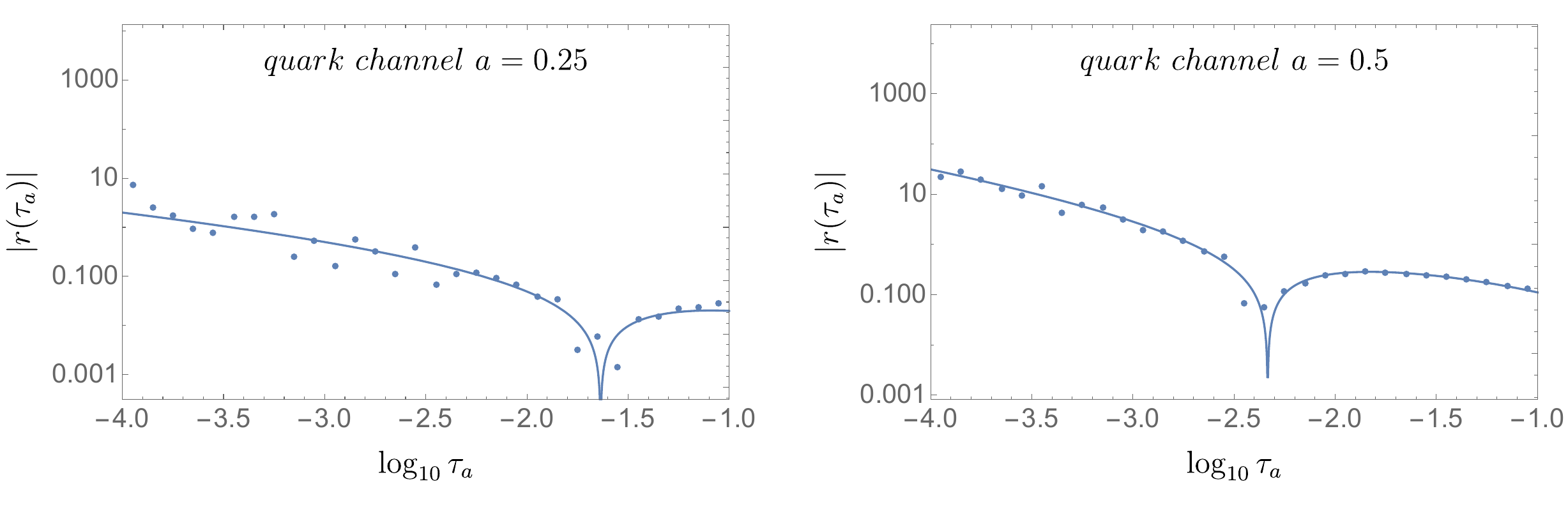}
\includegraphics[scale=0.33]{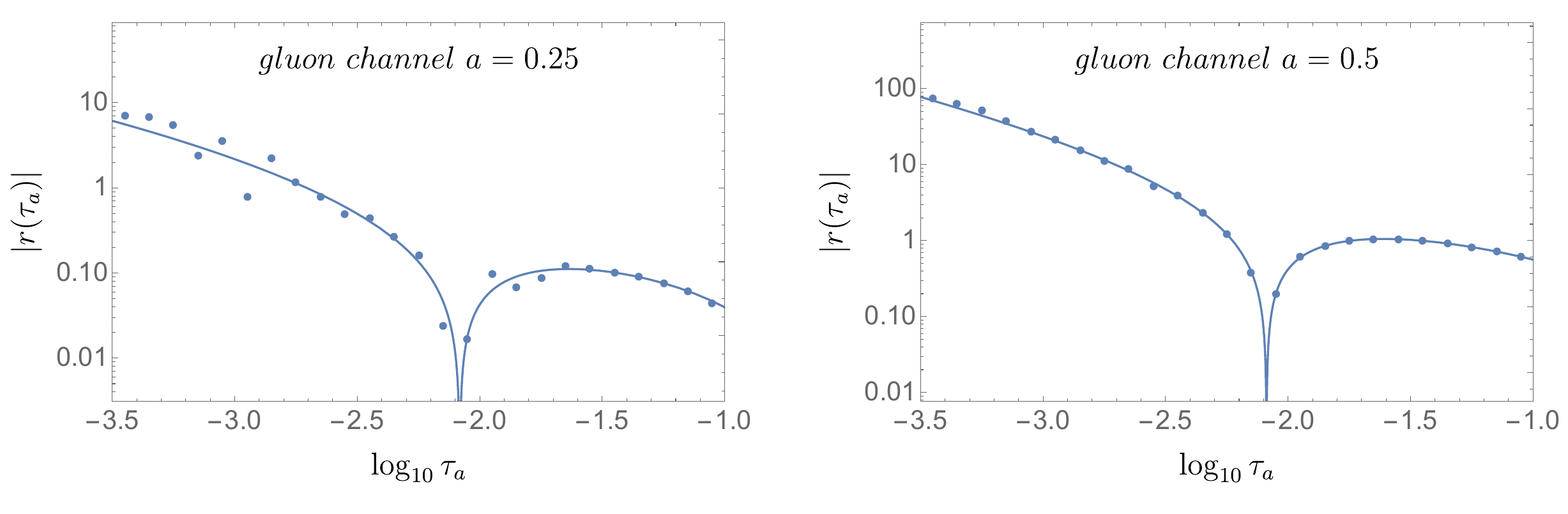}
    \caption{
   NLO nonsingular parts (dots) at $a=0.25$ and $0.5$ for quark (upper) and gluon (lower) channels and fit curves (blue) including recoil corrections}
\label{fig:recoil_nlo}
\end{figure}
\begin{multicols}{2}
\subsection{Two-loop constants}
\label{ssec:nlo}

Following the same strategy, we determine the two-loop constant of jet function by using NLO total rate \cite{Chetyrkin:1997iv}, NLO fixed-order obtained from MC simulations, and two-loop singular part obtained from \eq{fac1}. 

We also perform a fit to the nonsingular part by including recoil-correction terms for $a>0$. However, their scaling behavior at NLO is currently unknown. We make a conjecture for regarding the scaling behavior, test this conjecture in the quark channel, and then apply it to the gluon channel.
Our conjecture is that dominant scaling behavior can be inferred from crossing terms between the one-loop jet function and recoil corrections from the one-loop soft function. In this conjecture, we assume that recoil-corrections from the two-loop soft function and one- and two-loop jet functions will not produce more singular terms. 
The convolution of singular terms $\ln^k\ta$ from the jet function and recoil corrections of a form in \eq{fitRec} gives 
\begin{equation}\label{eq:rec}
\sum_{n=1}^{\lceil 1 / (1-a))\rceil-1} \sum_{k=0}^2 \,
\alpha_{n,k} \frac{\ln^k\ta}{\ta^{1-n (1-a)}}
\,,\end{equation}
where $\alpha_{n,k}$ are fit parameters. For the region $0<a\le 1/2$, $n=1$ and \eq{rec} reduces to
\begin{equation}\label{eq:rec1}
\sum_{k=0}^2 \, \alpha_{1,k} \frac{\ln^k\ta}{\ta^a}\,. \qquad (0<a\le 1/2)
\,.\end{equation}

We also need to include typical integer-power corrections of the form
\begin{equation}\label{eq:fitAsym2}
\sum_{k=0}^{3} \alpha_{k} \ln ^{k} \ta
\,,\end{equation}
where $\alpha_k$ are fit parameters. These terms are leading contributions for $a\le 0$ since there are no recoil corrections in \eq{rec}.
The form in \eq{fitAsym2} turns into $r_c+\ta \sum_{k=0}^3 \tilde{\alpha}_k  \ln^k\ta$ in the remainder function $r_c(\ta,1)$, and the value of $r_c$  is determined from the fitting to $r_c(\ta,1)$ data obtained from \eq{remainder}. 

A subtlety of the fit with recoil corrections in $a>0$ is that eight or more fit parameters make the fit ill-posed, and an ordinary chi-square fit cannot constrain those parameters properly. We adopt Tikhonov regularization \cite{tikhonov1963solution}, in which a regularization term $\lambda | \Gamma\mathbf{x}|^2$ is added to the chi-square and tames the ill-posed fit, where $\lambda$ is a regularization parameter and $\Ga$ is a Tikhonov matrix. More details about the regularization and our choice of $\lambda$ and $\Ga$ are discussed in \appx{Tikhonov}.

\fig{recoil_nlo} shows the nonsingular part of angularity distribution in the quark and gluon channel at $a=0.25$ and $0.5$ and, along with fitting results using terms in both \eqs{rec}{fitAsym2}. 
The fit curves well describe data. 
We also looked into the small $\ta$ region beyond the range in the figure and checked the relative importance of the leading-log term in \eq{rec}.
At $a=0.25$, the fit result with the leading-log is qualitatively similar to that with all three terms in \eq{rec}. 
However, at $a=0.5$, just leading-log cannot explain the data and including subleading-log terms in the fit form makes significant improvement. Therefore, we include all three terms from \eq{rec} in our fit.

Once $r_c$ is determined from the fit, one can obtained the two-loop constant $\cjt$ from the following relation 
\begin{equation}
r_c= -2\left(\frac{\as}{4\pi}\right)^2\cjt + \Delta_a
\,,\label{eq:rc2cj}
\end{equation}
where $\Delta_a=\Ga_t/\Ga_B-\left[A-2\left(\tfrac{\as}{4\pi}\right)^2\cjt\right]$ is obtained by rewriting \eq{tot}
and the term in square bracket is simply the sum of all contributions except for missing $\cjt$ hence, is known from the singular part.   
We determine the constants at seven values of $a=-1+n/4$, where $n=\{0,\cdots, 6\}$, for which numerical values of $\Delta_a$ for quark and gluon channels are 
\bea
\Delta_a^\text{quark} &=&\{-0.0301, -0.0273, -0.0243, -0.0207,\nn\\ \quad&&-0.0157, -0.0082, 0.0036\}
\nn\,,\\
\Delta_a^\text{gluon} &=& \{-0.1598, -0.1474, -0.1319, -0.1106,\nn\\ \quad&& -0.0787, -0.0258, 0.0695\}
\,,\eea
where we set $\as=0.1127$.

\subsubsection{quark channel}
\label{sssec:quark}

According to the procedure described in previous sections, we numerically obtain the remainder function $r_c(\ta,1)$ in the quark channel at seven values of angularity parameters from $10^{11}$ MC events  
computed on Intel Xeon Platinum 9242 CPU for about 3k CPU hrs 
with an IR cutoff $\alpha_\text{cut}=10^{-11}$ on a dimensionless variable $y_{ij,k}$ in the dipole subtraction \cite{Nagy:1998bb,Nagy:2003tz,Gleisberg:2007md}.

\begin{figure}[!htbp]
\centering
\includegraphics[scale=0.33]{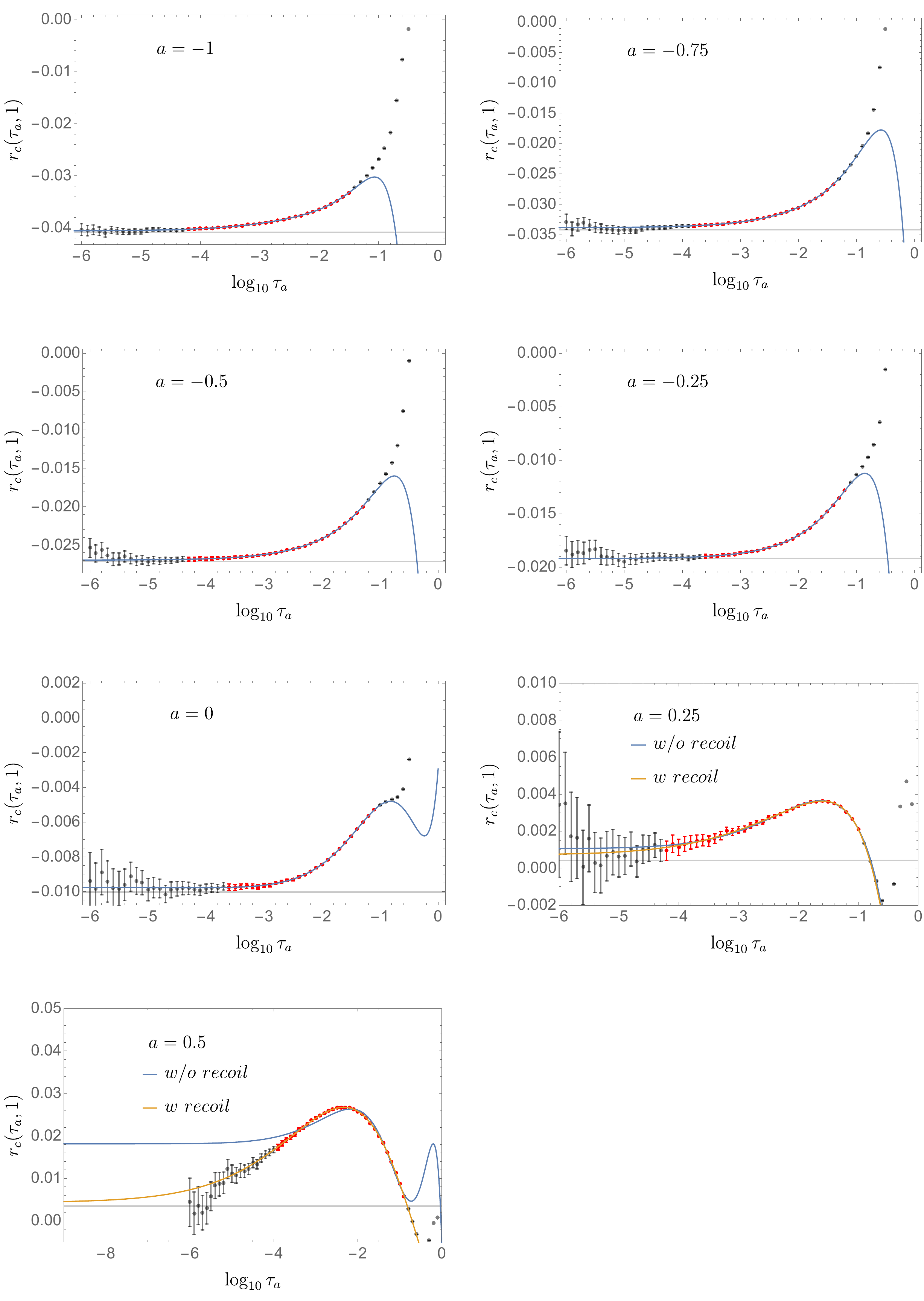}
\caption{
NLO remainder function $r_c(\ta,1)$ (gray and red dots with error bar) in the quark channel, and fit results without recoil corrections (blue) and with the corrections (orange) to the selected region of data (red dots with error bar) compared with existing values of $r_c$ (horizontal gray line)}
\label{fig:hqq_nlo}
\end{figure}

The remainder function in \fig{hqq_nlo} tends to have larger uncertainties in the small ${\ta}$ region around $10^{-4}$ and below, due to fine-tuning with large cancellations between fixed-order and singular parts.
In the plots for $a\le 0$, one can read plateau regions approaching existing results \cite{Bell:2018gce} (horizontal gray line) and determine $r_c$, hence two-loop constant $\cjt$ using \eq{rc2cj}.
The value of $r_c$ is determined by extrapolating the fit curve (blue) to $\ta=0$ with a fit form obtained by integrating \eq{fitAsym2}. In selection of the fit region (red dots with error bar), we eliminated the plateau region to test the asymptotic forms.

On the other hand, for $a>0$, a plateau is hardly observed from data, and two fit results with and without recoil corrections (orange and blue) are shown in the figure, respectively. Plateau region would appear in smaller $\ta$ region as $a$ increases because the recoil corrections like $\ta^{n(1-a)}$ make convergence of the remainder functions further slow. It is not accessible from MC results computed with double precision of machine variables. As indicated in \fig{recoil_nlo}, the fit with recoil corrections at $a=0.5$ clearly extrapolates to the value of $r_c$ close to existing value \cite{Brune:2022cgr,Bell:2021dpb} (horizontal gray line).
Our values of $\cjt$ are given in \tab{c2jqq} and show reasonable agreement with existing results in \cite{Bell:2018gce,Brune:2022cgr,Bell:2021dpb}.
We note that our fit region does not include the plateau region, and our results in \tab{c2jqq} tend would have larger uncertainties than those that could have been obtained from the best fit including the plateau region, because our purpose in the quark channel is to verify fitting with the asymptotic form.

In the fit region (red dots with error bar), we take a set of fit regions  $\{ \ta^\text{low}\,,\ta^\text{high} \}$ and obtain a set of $r_c$ values and statistical uncertainties from fitting to the regions. For the systematic uncertainty, which mainly comes from the IR cutoff and higher-power corrections, we take the maximum deviation of $r_c$ from its averaged value in the set. We obtain our central value from the averaged $r_c$, and our uncertainty from the sum of the systematic and averaged-statistical uncertainties in quadrature. 
Specifically, the set $\{ \ta^\text{low}\,,\ta^\text{high} \}$ is obtained from all possible combinations for pair from 9 succeeding points for $\log_{10} \ta^\text{low}$ domain, separated by a step size of $0.1$, and similarly 5 succeeding points for $\log_{10} \ta^\text{high}$ domain. Our selection criteria for $\ta^\text{high}$ domain is that it should be robust against higher-power corrections and we select $\{-1.6+0.4(1+a) , -1.2+0.4(1+a)\}$. The criteria for $\ta^\text{low}$ domain is that it should be insensitive to IR cutoff. We scan all available domains of 9 succeeding points and select the domain giving least that gives the least standard deviation for $r_c$. We also explore the effect of the cutoff $\alpha_\text{cut}$ by comparing our standard value $\alpha_\text{cut}=10^{-11}$ with larger value $10^{-8}$ in \appx{cutoff}

\end{multicols}
\begin{table}
\centering
\begin{threeparttable}[!htb]
\caption{Two-loop constant $\cjt$ of quark-jet function for angularity} \label{tab:c2jqq}
\begin{tabular}{cccccccc}
\toprule
$\cjt\,\textbackslash\, a$         & -1       & -0.75    & -0.5     & -0.25    & 0            \\ 
\midrule
this work & $65.43\pm0.91$ & $40.41\pm1.02$ & $16.52\pm0.36$ & $-9.10\pm1.16$ & $-36.84\pm0.72$  \\

Ref.~\cite{Bell:2018gce} & $66^{-5.2}_{-3.5}$&$42.3_{-3.3}^{+5.1}$&$17.3_{-2.5}^{+3.2}$&$-9.34_{-2.48}^{+2.76}$&$-36.3_{-2.4}^{+2.7}$ \\
Ref.~\cite{Brune:2022cgr}\tnote{*} & 67.8 & 43.1 & 18.2 & $-8.7$ & $-34.9$ \\
\midrule
&0.25  &0.5 \\
\midrule
&$-55.36\pm0.69$& $-3.18\pm2.13$\\
&$-57.6_{-3.2}^{+3.8}$&$-79.8^{+39.7}_{-24.9}$ \\
& $-53.6$ & $-0.22$ \\
\bottomrule
\end{tabular}
\begin{tablenotes}
     \item[*] The values are read from plots in the thesis and would have errors of a few percent.
   \end{tablenotes}
  \end{threeparttable}
\end{table}
\begin{multicols}{2}

\subsubsection{gluon channel}
\label{sssec:gluon}
Now we determine the two-loop constant of the angularity gluon-jet function and compare our results to values given in \cite{Brune:2022cgr}.
The procedure is essentially the same as that described in \sssec{quark}, and we only describe the difference from that section. 

\begin{figure}[htbp]
\centering
\includegraphics[scale=0.33]{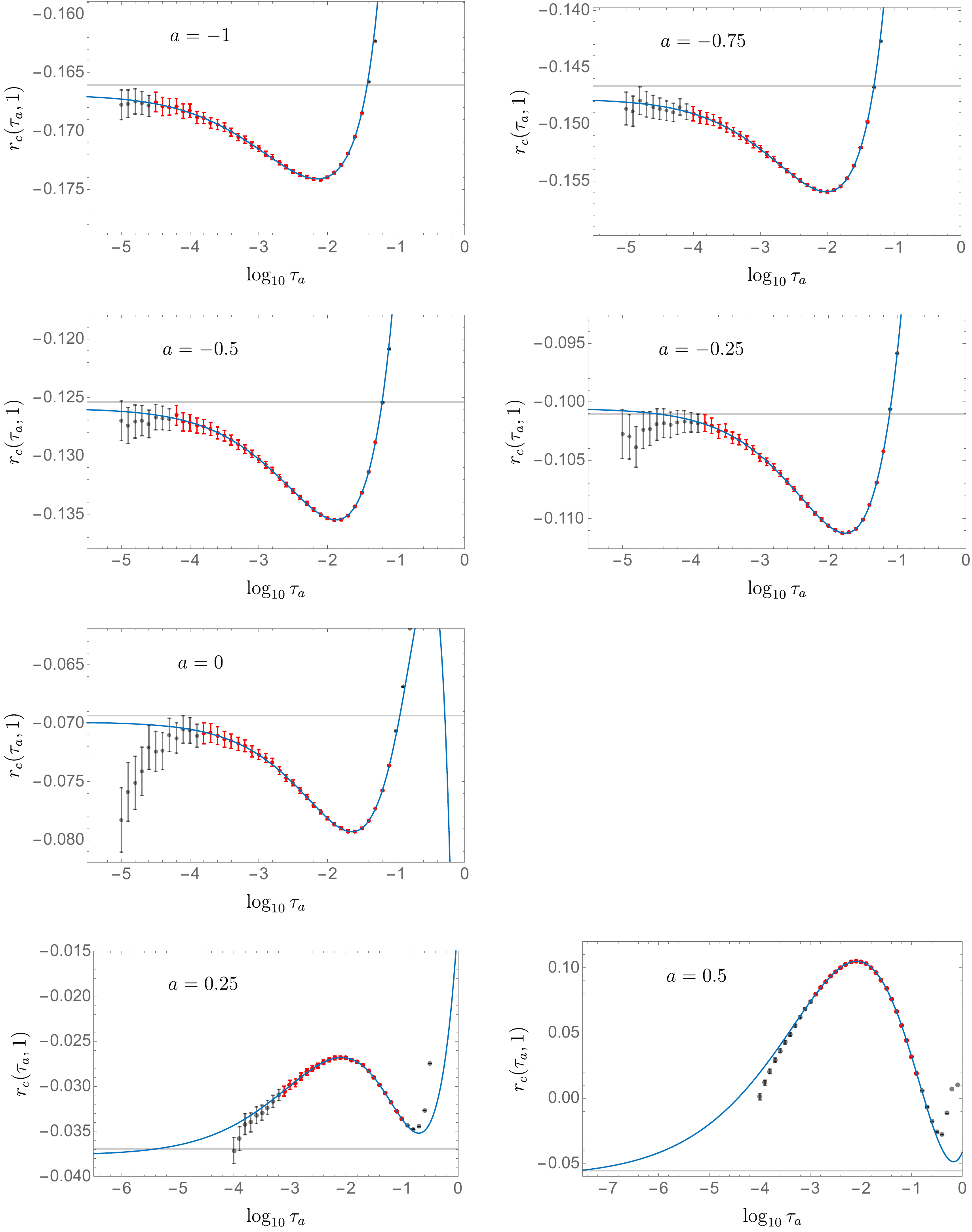}
\caption{NLO remainder function $r_c(\ta,1)$ (gray and red dots with error bar) in the gluon channel, and fit results (blue) from the selected region (red dots with error bar) compared with existing values of $r_c$ (horizontal gray line)}
\label{fig:hgg_nlo}
\end{figure}

We use the same number of $10^{11}$ MC events with larger cutoff $\alpha_\text{cut}=10^{-8}$ in the gluon channel. \fig{hgg_nlo} shows NLO remainder function $r_c(\ta,1)$ (gray and red dots with error bar) obtained from MC events, compared with fit results (blue) for the selected region of data (red dots) and $r_c$ values in \cite{Brune:2022cgr} (horizontal gray line). 
The fits for $a\le 0$ were performed without the recoil corrections while for $a>0$ with the recoil corrections.
Unlike the quark channel, we do not observe any plateau regions at any value of $a$ hence, fitting with asymptotic form in \eqs{rec}{fitAsym2} is essential in the gluon channel.

Due to more severe cutoff effect, the fit regions are narrowed.
The domains of $\log_{10} \ta^\text{low}$ reduces to 7 succeeding points for non-positive $a$ and $\log_{10} \ta^\text{high}$ to 5 points for positive $a$. 

The cutoff effect on the value of $\cjt$ in the gluon channel is shown in \appx{cutoff}. \tab{c2jgg} shows our two-loop constant $\cjt$ converted from $r_c$ values by using \eq{rc2cj}, and our values show reasonable agreement with those values reported in \cite{Brune:2022cgr}.
\end{multicols}
\begin{table}[H]
\centering
\begin{threeparttable}
\caption{two-loop constant $\cjt$ of gluon-jet function for angularity} \label{tab:c2jgg}
\begin{tabular}{cccccccc}
\toprule
$\cjt \,\textbackslash\, a$         & -1       & -0.75    & -0.5     & -0.25    & 0            \\ 
\midrule
this work & $44.19\pm1.70$ & $2.10\pm2.75$ & $-36.43\pm1.65$ & $-62.08\pm2.49$ & $-54.55\pm2.60$  \\
Ref.~\cite{Brune:2022cgr}\tnote{*}  & $37.18$&$-4.59$&$-40.27$&$-56.95$&$-55.73$ \\
\midrule
$\cjt \,\textbackslash\, a$    &0.25  &0.5 \\
\midrule
this work &$75.09\pm10.99$& $814.48\pm28.71$\\
Ref.~\cite{Brune:2022cgr}\tnote{*} &$69.21$&$776.61$ \\
\bottomrule
\end{tabular}
\begin{tablenotes}
     \item[*] The values are read from plots in the thesis and would have errors of a few percent.
   \end{tablenotes}
\end{threeparttable}
\end{table}
\begin{multicols}{2}

\section{Angularity distributions at \NNLLp matched to NLO}
\label{sec:angdis}
In this section, we present the angularity distribution at \NNLLp matched to fixed-order NLO results, which is $\mathcal{O}(\as^2)$ correction relative to the Born rate $\Ga_B(\mu)$.
The distribution is composed of two parts
\be 
\frac{1}{\Ga_B}\frac{d \Ga}{d \ta}
=\frac{1}{\Ga_B}\frac{d \Ga^\text{res}}{d \ta}
+r(\tau_a,\mu_{\text{ns}})
 \,,
\label{eq:sigta1}
\ee
where $\tfrac{d \Ga^\text{res}}{d \ta}$ is the resummed-singular part, and $r(\tau_a,\mu_{\text{ns}})$ is the nonsingular part at a scale $\muns$, including all subleading-power corrections of $\ta$ at NLO. 
As described in \sec{form}, we perform the log resummation by RG-evolving the functions in factorization equation \eq{fac2}. We give a final expression in \eq{sigta2}, and the detailed procedure is summarized in \appx{sscet} and \appx{res}. The nonsingular part makes our full distribution in \eq{sigta1} matched to fixed-order result as defined in \eq{fqcddiff}. 
Including this is important because the unphysical behavior of singular terms in the large $\ta$ region can be corrected by adding the nonsingular part.
Note that in \eq{sigta1}, $\Ga_B=\Ga_B(m_H)$ is the Born decay rate at the scale of Higgs mass hence, both the singular and nonsingular parts contain a prefactor $\Ga_B(\mu)/\Ga_B(m_H)=\as(\mu)^2/\as(m_H)^2$.

 We also compute a cumulative distribution by integrating \eq{sigta1} from 0 to $\ta$ as

\begin{equation}
    \begin{aligned}
      \label{eq:cum}
\frac{\Ga_c\left(\ta\right)}{\Ga_B}
&= \int_0^{\ta} d \ta^{\prime}  \left[\frac{1}{\Ga_B}\frac{d \Ga^\text{res}}{d \ta^{\prime}}+r(\ta')\right]
\\&=\frac{\ta}{\Omega\,\Ga_B} \frac{d \Ga^\text{res}}{d \ta}+r_c(0,\ta)
\,,  
    \end{aligned}
\end{equation}
where $\Omega=\eta_S+2\eta_J$ is the sum of integrated anomalous dimensions of the jet and soft functions, which are defined in \eqs{Keta-def}{ker}. When the resummation is turned off, $\Omega$ reduces to zero and cancels against the same factor in $d \Ga^\text{res}/d\ta$ so that $\Ga_c (\ta)$ remains finite.

\subsection{Matching}
\label{ssec:nonsing}

The nonsingular part in \eq{sigta1} can be expressed in a scale independent form satisfying $ dr\left(\tau_a, \mu\right)/d\mu=0$ as  
\begin{equation}
    \begin{aligned}
   \label{eq:ns}
 r&\left(\tau_a, \muns\right) =\frac{\as(\muns)^2}{\as(m_H)^2}
 \bigg\{
 \frac{\alpha_s\left(\muns\right)}{4 \pi} r^1\left(\tau_a\right) 
\\ &\quad+\left(\frac{\alpha_s\left(\muns\right)}{4 \pi}\right)^2\left[r^2\left(\tau_a\right)+3\beta_0 r^1\left(\tau_a\right) \ln \frac{\mu^2_{\mathrm{ns}}}{m_H^2}\right]
\bigg\}
  \,,     
    \end{aligned}
\end{equation}
where $r^1(\ta)$ and $r^2(\ta)$ are LO and NLO coefficients, respectively and only depend on $\ta$. The scale dependency on $\muns$ is made explicit with the logarithmic term proportional $3\beta_0 r^1$, which is cancelled by the same term generated by scale variation of lower-order term proportional to $r^1(\ta) $. In comparison to nonsingular part in $e^+e^-$  angularity \cite{Bell:2018gce}, the log term differs from that in \cite{Bell:2018gce} by a factor of 3 due to an additional prefactor $\as(\muns)^2$ from the Born rate $\Ga_B^g(\mu)$, which is absent in $e^+e^-$ angularity.

An analytic expression of the nonsingular part is unknown except for the thrust limit $a=0$, and it is determined numerically from MC events obtained in \ssec{nlo}.
We interpolate the remainder function obtained from MC events instead of interpolating the nonsingular part 
because the binned-nonsingular part inherits uncertainties by an amount of bin size in x-axis associated while with its integration, remainder function, does not at the end of each bin. Then, we obtain the nonsingular part by taking the numerical differentiation of the interpolated remainder function. 

In order to maintain good precision of the interpolation in both small and large $\ta$ regions, the spectrum of $r^n(\ta)$ is divided into two regions, below and above a point $\ta=0.1$. Below the point, MC data is binned in log space from a lower bound $\log_{10} \ta^\text{low}$ to the point with a step size of $0.1$, where the lower bound is $-7$ at LO and $-4$ or, $-5$ at NLO.
Above the point, the MC data is binned in linear space $\ta$ up to $1$ with a step size of $0.01$. 
At NLO, MC data in small $\ta$ regions tends to be significantly biased by the cutoff effect as shown in \fig{yijk}, and the fit function is instead to describe points below $\ta=0.1$.
Then, interpolation functions $r_{\log_{10} }^n(\ta)$ and $r_{\operatorname{lin}}^n (\ta )$ in two regions are joined smoothly following the method in \cite{Bell:2018gce}:
\begin{equation}
    \begin{aligned}
     \label{eq:ns_lojoin}
r^n\left(\tau_a\right)&=\left[1-f\left(\tau_a, 0.1,0.01\right)\right] r_{\log_{10} }^n\left(\tau_a\right)\\&+f\left(\tau_a, 0.1,0.01\right) r_{\operatorname{lin}}^n\left(\tau_a\right)
\,,  
    \end{aligned}
\end{equation}
where $f\left(z, z_0, \epsilon\right)=1 /\left(1+e^{-\left(z-z_0\right) / \epsilon}\right)$ is a transition function.

\begin{figure}[H]
\centering
\includegraphics[scale=0.31]{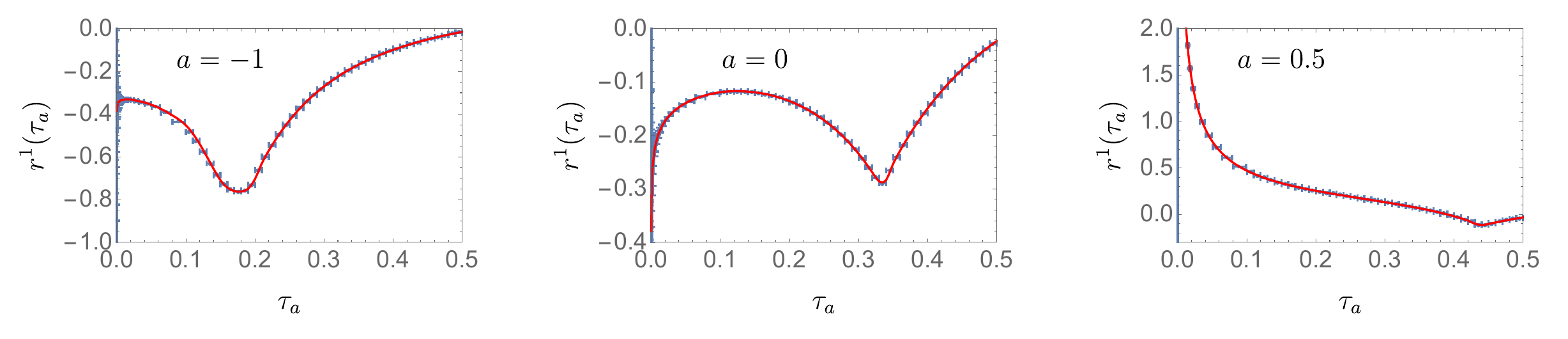}
\includegraphics[scale=0.305]{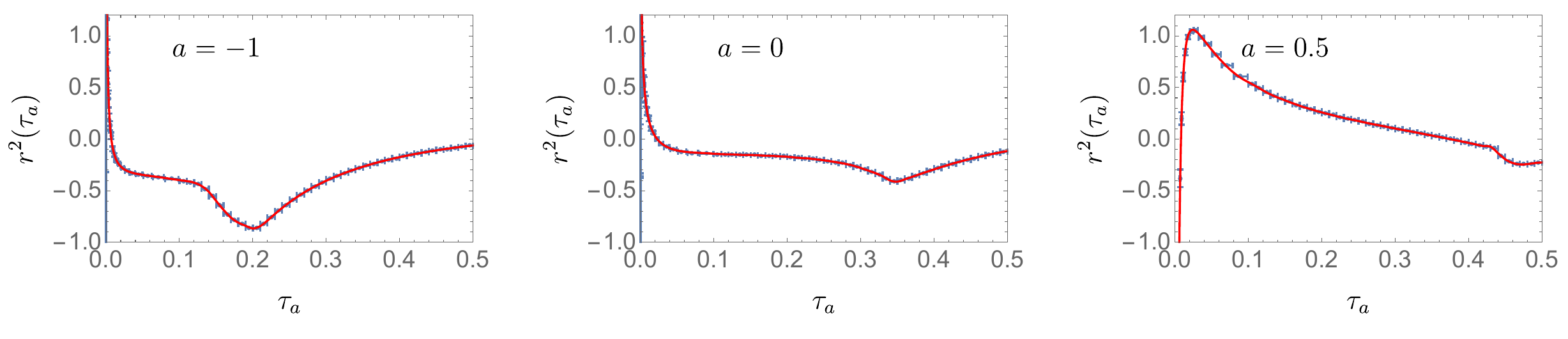}
\caption{LO (upper) and NLO (lower) coefficients of nonsingular part  $ r\left(\tau_a, \muns\right)$ at $a=-1,\, 0,\, 0.5$.}
\label{fig:ns_lo}
\end{figure}

Nonsingular distributions with error bars at LO (first row) and at NLO (second row) are shown in \fig{ns_lo}, and red curves are the interpolations joined using \eq{ns_lojoin}.  Note that the nonsingular is not a physical quantity and can be negative, while full distribution in \eq{sigta1} is physical and positive definite. 

\end{multicols}
\begin{multicols}{2}

\subsection{Numerical results}
\label{ssec:num}

For the numerical calculation of distribution, one needs to make proper choice for the scales: hard scales $\mu_{C_t}$, $\mu_{C_S}$, soft scale $\mu_S$ and jet scale $\mu_J$. 
In the choice one needs to consider the $\ta$ distribution divided by three regions: the peak region $\left(\tau_{a} \sim 2 \Lambda_{QCD} / m_H \ll 1\right)$, the tail region $\left(2 \Lambda_{QCD} /m_H \ll \tau_{a} \ll 1\right)$ and the far-tail region $(\ta\sim 1)$. 
The tail region is the region where $\ta$ remains small and resummation is effective. In the region, the natural choice is the canonical scales $\mu_{C_t,C_S}= m_H,\, \mu_{J} = m_H \ta^{1/(2-a)},\,  \mu_S = m_H \ta$ that minimize the logarithms in each function. 
In far-tail region, $\mu_i \!\sim\! \mathcal{O}(m_H)$ and we can set all scales to be hard scale $\mu_i=m_H$, which is a conventional choice in fixed-order perturbation theory.  
The peak region is dominated by nonperturbative (NP) effect and our perturbative approach is invalid. Before approaching the peak region, we need to stop scales running into $\Lambda_{QCD}$ and make them frozen in perturbative region well above $\Lambda_{QCD}$.
One can design \textit{Profile function} $\mu_i(\ta)$ that is a smooth function of $\ta$ satisfying above criteria \cite{Abbate:2010xh,Ligeti:2008ac,Berger:2010xi}. 
Here, we adopt the profile functions and scale variations to estimate perturbative uncertainty designed for the $e^+e^-$ angularity in \cite{Bell:2018gce} and we modify a parameter $t_2(a)$ to be $0.43\times 0.674^{1 - 1.913a}$, which is the value of $\ta$ where singular and nonsingular parts of the gluon channel are the same in size and beyond the point all scales smoothly transit to the hard scale. We also take the choice of \cite{Bell:2018gce} for a scale $\muns$ in nonsingular part.
\end{multicols}
\begin{figure}[!htbp]
\centering
\includegraphics[scale=0.37]{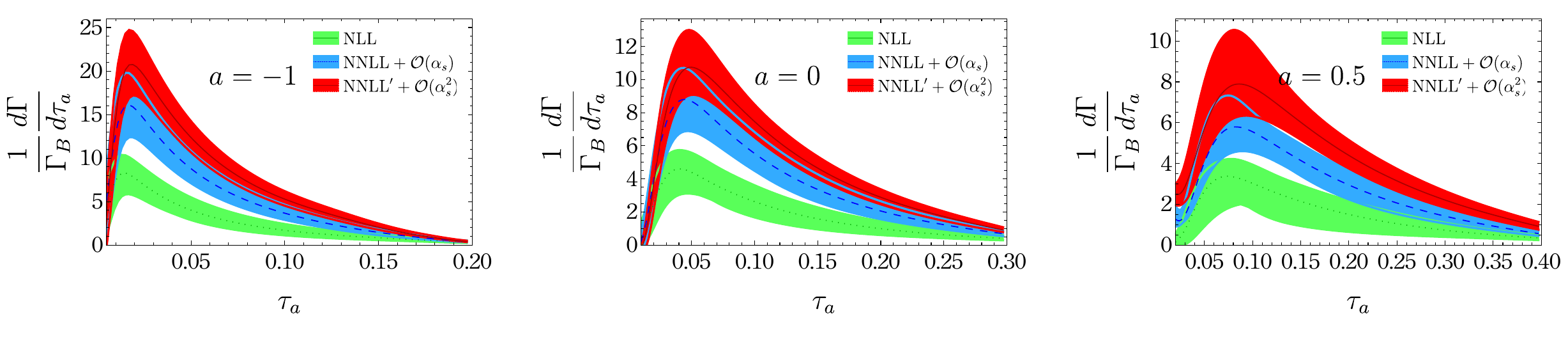}
\includegraphics[scale=0.37]{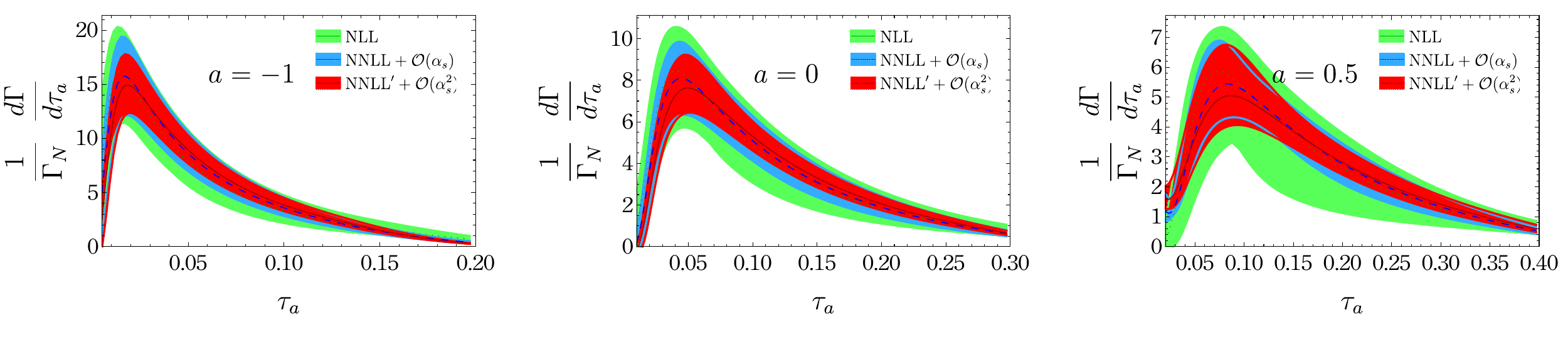}
\includegraphics[scale=0.37]{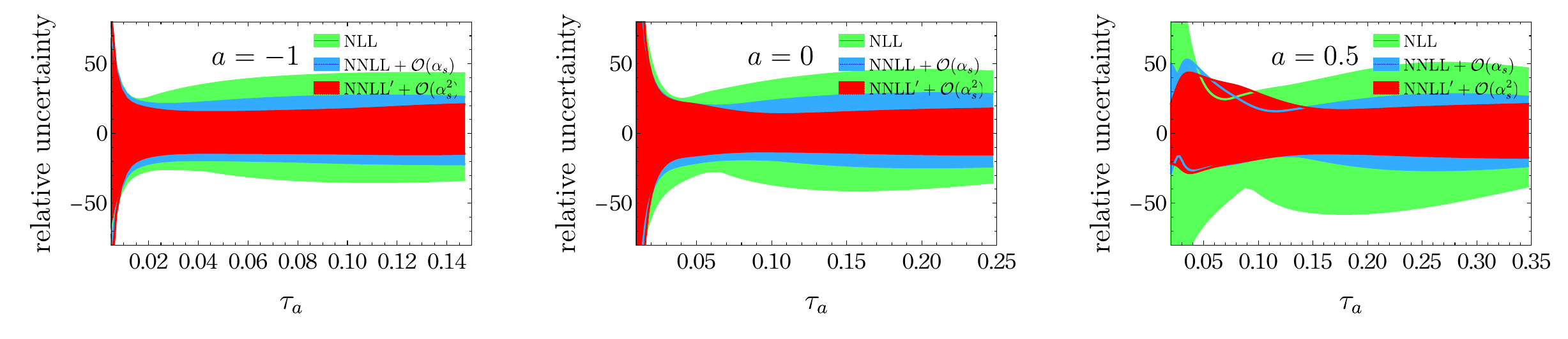}
\includegraphics[scale=0.37]{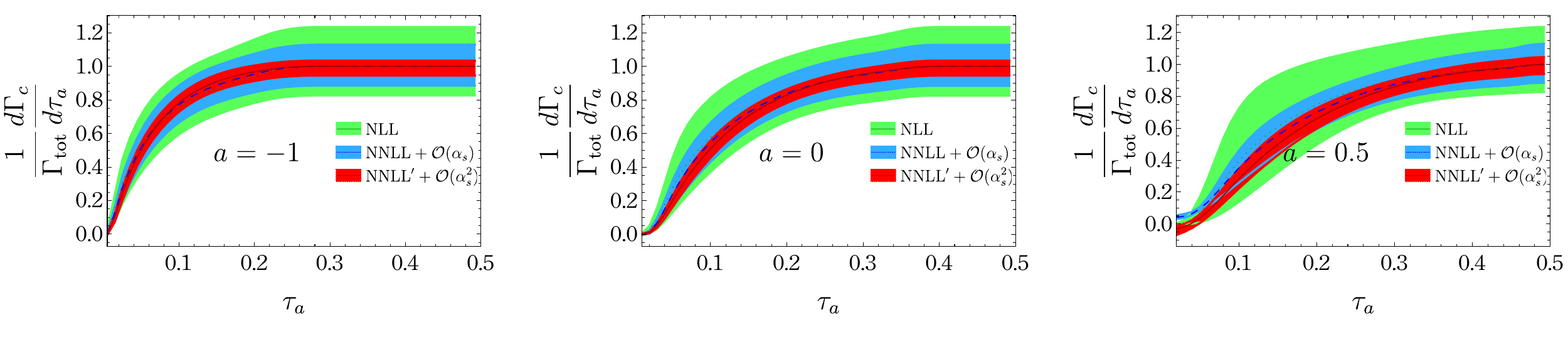}
\caption{First and second rows: Angularity distributions at three values of $a=-1,\,0,\,0.5$ for gluon channel in Higgs decay. The bands indicate perturbative uncertainties. The first row is normalized by the Born rate $\Gamma_B$ and the second by area covered by resummed distribution $\Gamma_N$ at corresponding order. 
Third row: Relative perturbative uncertainty in percentage. Bottom row: Cumulative distributions normalized by total rate.}
\label{fig:resum}
\end{figure}
\begin{multicols}{2}

\fig{resum} shows our resummed results at \NNLLp$+\mathcal{O}(\as^2)$ accuracy (red) for angularity distribution (first and second rows), relative uncertainties (third row), and cumulative distributions (fourth row) at three values of $a=-1$, 0, and 0.5 and results at lower accuracy: NLL (green) and NNLL$+O(\as)$ (blue) for a comparison. Band at each accuracy represents perturbative uncertainty estimated by scale variations in the profile function.
The distributions normalized by the Born rate in the first row hardly overlap between different accuracy, and the peak of distribution increases by $100~\%$ from NLL to NNLL and $50~\%$ from NNLL to \NNLLp because of large perturbative corrections as in the total rate.
The distributions normalized by their own area in the second row show reasonable overlapping within the uncertainties between different accuracy which indicates a perturbative convergence of the resummation.
In the third row, the perturbative uncertainties with respect to central values reduce to about $20~\%$ at \NNLLp, which is about half of  NLL uncertainties, however the reduction rate with increasing accuracy is rather slow compared to the quark channel as many gluon initiated processes do. 
$\cjt$ uncertainties  in \tab{c2jgg} propagate to the distribution and the values also show deviations from 2-loop results in Ref.~\cite{Brune:2022cgr}, their effect in the distribution are as large as 0.8\%, which is negligible compared to the perturbative uncertainties at this level of accuracy.
In the fourth row, the cumulative distribution is normalized by the total rate at corresponding order in $\as$. 
One can observe the cumulative distribution reduces to the total rate in the far-tail region. In the region resummation is turned off and resummed distribution in \eq{sigta1} reduces to the fixed-order result in \eq{fqcddiff}.

One can observe a generic feature of the distribution that the shape becomes broad and the peak moves to right with increasing $a$ because for a given jet scale $\mu_J=m_H\ta^{1/(2-a)}$, increasing $a$ scales up the value of $\ta$. But this is not true for the soft scale $\mu_S=m_H \ta$, where $a$ does not change the scale of $\ta$. Both of them affect the distribution and make its change as a function of $a$ nontrivial. 
The peak of the distribution is essential feature of the resummation known as Sudakov exponent, which is $e^{-\frac{2\as C_A }{\pi}\ln^2 \ta}$ at LL accuracy, and cures the singular behavior $\ln \ta/ \ta$ and the peak position of gluon channel is rather located at large $\ta$ compared to that in the quark channel due to the Casimir scaling.

\begin{figure}[tbhp]
\centering
\includegraphics[scale=0.36]{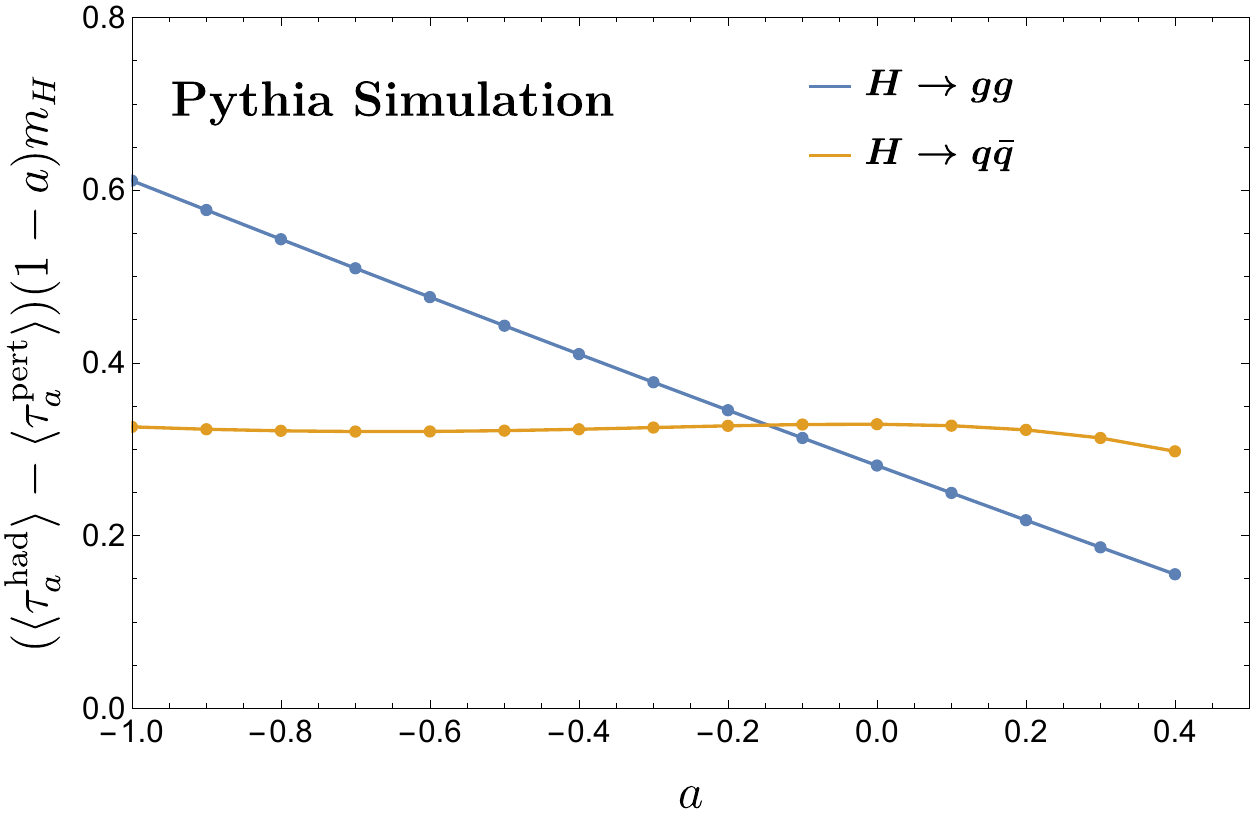}
\caption{Difference between $\tmean{a}{had}$ at hadron level and $\tmean{a}{pert}$ at parton level from Pythia simulation for gluon channel (blue) and for quark channel (orange).  }
\label{fig:deltahp}
\end{figure}

\fig{resum} shows pure perturbative contribution without the NP effect like hadronization that can be included by introducing a NP model \cite{Ligeti:2008ac,Kang:2013nha,Kang:2014qba} or, power correction with a universal NP parameter $\Omega \sim \cO(\Lqcd)$ \cite{Lee:2006nr}, which can be further refined with a scheme subtracting renormalon ambiguities \cite{Hoang:2007vb,Hoang:2008fs}. 
The parameter $\Omega$ captures a leading NP correction and shifts the perturbative distribution, hence the first moment of $\ta$ as 
\be
\tmean{a}{had}= \tmean{a}{pert} +\frac{2}{1-a}\frac{\Omega}{m_H}+\omgll
\,,\label{eq:tamean}
\ee
where the superscript `had' is for the hadron level that can be measured in experiment, or simulated in event generators, and `pert' is for pure perturbative result at the parton level. 
\eq{tamean} includes the quark and gluon channel $\Omega=f_q\Omega_q+f_g \Omega_g$, where  $f_{i}=\Ga^{i}_{tot}/(\Ga^q_{tot}+\Ga^g_{tot})\approx 0.9$ and $0.1$ for the quark and gluon fractions, respectively.  
For the quark channel $\Omega_q^{ee}\approx 0.35$~GeV from the $e^+e^-$ thrust analysis \cite{Abbate:2010xh} and assuming Casimir scaling estimates $\Omega_g\approx C_A/C_F\Omega_q$. 
We can determine $\Omega_g$ in Higgs factories by using the known value of $\Omega_q^{ee}$.

\eq{tamean} predicts $a$ dependence $(1-a)^{-1}$ in $\Omega$ term that can be easily identified by comparing experiment and theory predictions in $a$ space \cite{Berger:2003pk}. 
We can also test if the $a$ dependence in $\Omega$ term is comparable with the hadronization model in event generators.
\fig{deltahp} shows the difference $\tmean{a}{had}-\tmean{a}{pert}$ with a weight $(1-a)m_H$ obtained from Pythia simulations for the quark channel and gluon channel. The $y$ values correspond to $2\Omega_q$ for the quark and $2\Omega_g$ for the gluon and would weakly depend on $a$ via subleading term $\omgll$ in \eq{tamean}. While the quark channel is consistent with the $a$ dependence and $\Omega_q^\text{pythia}\approx 0.2$~GeV roughly agreeing with $\Omega_q^{ee}$ within a factor of two, the gluon channel shows rather stronger $a$ dependence ranging $\Omega_g^\text{pythia}\approx 0.2\sim 0.6$~GeV  order of $\Lqcd$. 
Their relative contributions are $f_q\Omega_q^\text{pythia}\approx 0.2$ and  $f_g \Omega_g^ \text{pythia}\approx 0.02\sim \! 0.06$, where the gluon contributions are suppressed and the $a$ dependence would become soft. Measurements in future Higgs factories can justify the value of $\Omega_g$.
If Pythia value is different from the measurement, it could come from another source such as large perturbative corrections at higher orders absent in Pythia that have been absorbed into hadronization parameters fitted to experimental data. 

\begin{figure}[tbhp]
\centering
\includegraphics[scale=0.40]{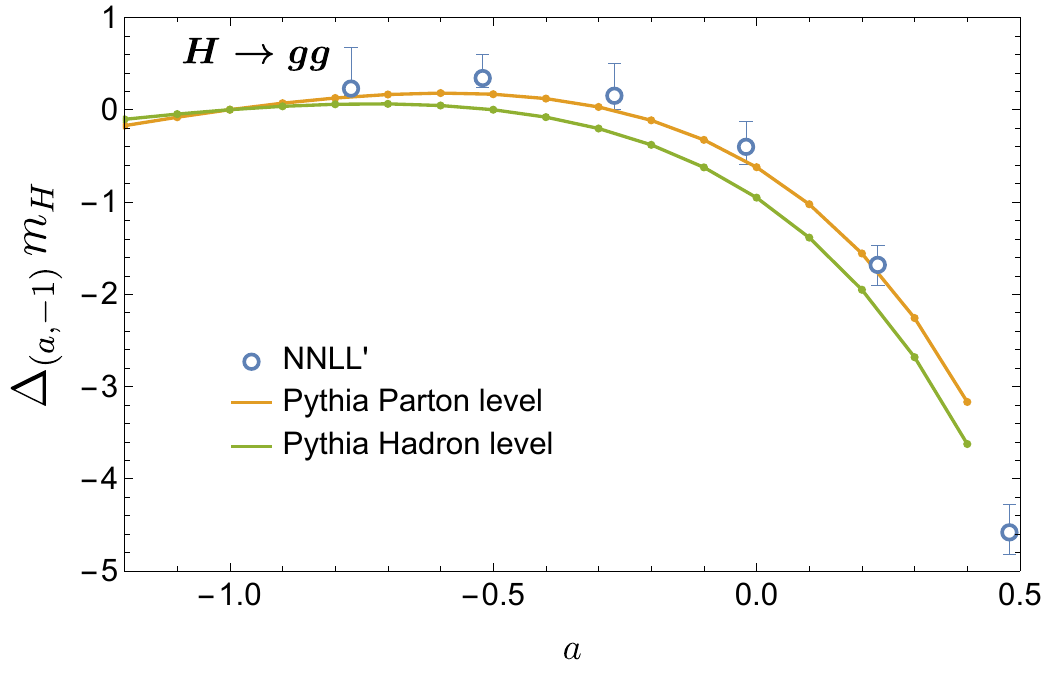}
\includegraphics[scale=0.40]{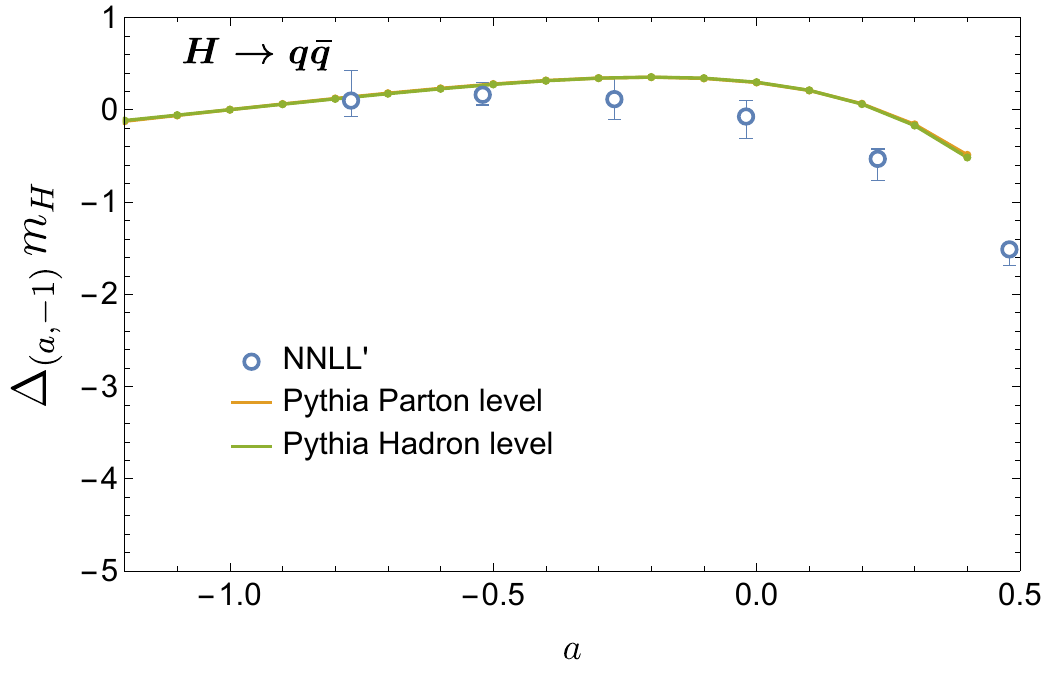}
\caption{Prediction of subtracted moment $\Delta_{(a,-1)}$ as a function of $a$ from resummed results at \NNLLp compared with Pythia simulations at parton and hadron levels for the quark channel (left) and for the gluon channel (right). }
\label{fig:deltaqg}
\end{figure}

The subleading NP corrections of $\mathcal{O}(\Lqcd^2/m_H^2)$ in \eq{tamean} can be studied from higher-moment or cumulant analysis.
In $e^+e^-$ thrust \cite{Abbate:2012jh}, 2nd thrust cumulant, which is completely insensitive to $\Omega^q$ enabled to determine a subleading correction $\Omega_2^{'q}(0)$, which is the sum of two contributions: one from leading thrust factorization and the other from a power correction $\Omega^q_{1,1}(0)\sim \Lqcd^2$ multiplied by a perturbative part. This can be extended to $e^+e^-$ angularity to determine $\Omega_2^{'q}(a)$ at nonzero $a$ and to angularity in Higgs decay to determine corresponding corrections in the gluon channel.
Following the convention of \cite{Abbate:2012jh}, the correction $\mathcal{O}(\Lqcd^2/m_H^2)$  in \eq{tamean} can be parameterized as $2M_{0,1}(a) \,\Omega_{1,1}(a)/m_H^2$, where $\Omega_{1,1}(a)$ is the correction appearing in the 2nd cumulant and  $M_{0,1}(a)$ accounts for a perturbative part. 
 Another way to study the subleading NP corrections is to take the difference between moments as in \cite{Chien:2018lmv,Kang:2018agv} at two points in $a$ space
\begin{equation}
\begin{aligned}
   \Delta_{(a,a')}
&=(1-a)\tmean{a}{had} -(1-a')\tmean{a'}{had}
\\
&=(1-a)\tmean{a}{pert} -(1-a')\tmean{a'}{pert}+\omgll
\,,
\label{eq:delaa}      
\end{aligned}
\end{equation} 
where the leading correction $\Omega$ is cancelled in the second line and the subtracted moment $\Delta_{(a,a')}$ is sensitive to the subleading correction that can be expressed as a difference of the quantity $2(1-a)M_{0,1}(a) \,\Omega_{1,1}(a)/m_H^2$ at two points $a$ and $a'$. By comparing experimental measurement and perturbative predictions one can determine the correction. The subtracted moment with angularity provides additional information different from that obtained from thrust cumulant analysis hence, is a useful tool for extracting information on subleading NP corrections.   
\fig{deltaqg} shows our perturbative predictions of \eq{delaa} at \NNLLp accuracies compared with Pythia simulations at parton and hadron levels for the gluon channel (left) and the quark channel (right). The Pythia results in the gluon channel show a notable difference between parton and hadron levels, while the difference in the quark channel is very small. 
This analysis can be done for different colliders such as Electron-Ion Collider \cite{AbdulKhalek:2021gbh} with DIS trust and angularity \cite{Kang:2013nha,Kang:2014qba,Chu:2022jgs,Zhu:2021xjn} and determine NP corrections. 
\end{multicols}
\begin{multicols}{2}

\section{Conclusion}
\label{sec:con}
We present improved predictions of angularity distribution $\ta$ in hadronic decays of Higgs boson via effective operator $H\to gg$ that suffers from large perturbative uncertainties compared to the other contribution via Yukawa interaction $H\to q\bar q$. 
The distribution is improved by resumming large logarithms of angularity at \NNLLp accuracy in the framework of SCET and by matching the resummed results to the fixed-order result at NLO. 

In order to achieve \NNLLp accuracy, we independently determined a remaining ingredient, 2-loop constant of the gluon-jet function for angularity. We find our values of the constant show reasonable agreement with the values in the literature.
In the determination of the constant, there is significant contamination from subleading singular corrections slowly suppressed in the small $\ta$ limit.
In order to estimate the correction, we use asymptotic forms in the small $\ta$ region and fit it to the nonsingular part of fixed-order result.
The asymptotic form for $a\le 0$ is $\sum_{k}\alpha_k \ln^k \ta$, which is one power-suppressed by the singular part, and for $0<a\le 1/2$ it has additional terms with fractional power $\sum_{k}\alpha_{1,k} \ln^k \ta/(\ta)^{a}$, which is conjectured from recoil corrections in one-loop soft function. The form in $a> 0$ has eight degrees of freedom and the fitting result is sensitive to noise hence, ill-posed.
We found that it can be handled by adopting Tikhonov regularization that reduces the sensitivity to noise.

We also consider nonperturbative corrections in the angularity. 
The corrections receive two contributions, one from the quark channel and the other from the gluon channel, and by using the value known for the quark in $e^+e^-$ experiment, the gluon contribution can be determined from measurement in Higgs factories. Pythia simulation shows that the quark channel is comparable to the prediction of leading-power correction for the angularity, while the gluon channel is quite different from the prediction in $a$ dependence and magnitude estimated from Casimir scaling.
We also considered subtracted moment $\Delta_{(a,a')}$ defined by the difference between moments of $\ta$ at $a$ and $a'$ that mitigates the leading correction hence, it is sensitive to subleading corrections. The subtracted moment as a function of $a$ is an independent direction from the cumulant analysis and provides additional information to the subleading correction. 

\appendix
\end{multicols}
\section{SCET functions}\label{app:sscet}

Here, we summarize the functions appearing in the factorization in the gluon channel, and the tag $g$ representing the gluon is implied implicitly. In Laplace space each function $G=\{C_t,C_S,\tilde{J},\tilde{S}\}$ in \eq{fac2} can be expressed as:

\be\label{eq:str}
\begin{aligned}
G\left(L_{G}, \mu_G\right)=& \sum_{n=0}^{\infty}\left(\frac{\alpha_{s}\left(\mu_{G}\right)}{4 \pi}\right)^{n} G^{(n)}\left(L_{G}\right) \\
G^{(0)}\left(L_{G}\right) =& 1
\\
G^{(1)}\left(L_{G}\right) =&
\left[\Ga_{G}^{0} L_{G}^{2}-\gamma_{G}^{0} L_{G}+c_G^{1}\right] \\
G^{(2)}\left(L_{G}\right) =&
\left[\frac{1}{2}\left(\Ga_{G}^{0}\right)^{2} L_{G}^{4}-\Ga_{G}^{0}\left(\gamma_{G}^{0}+\frac{2}{3} \beta_{0}\right) L_{G}^{3}\right.\\
&\left.+\left(\Ga_{G}^{1}+\frac{1}{2}\left(\gamma_{G}^{0}\right)^{2}+\gamma_{G}^{0} \beta_{0}+c_{G}^{1} \Ga_{G}^{0}\right) L_{G}^{2}-\left(\gamma_{G}^{1}+c_{G}^{1} \gamma_{G}^{0}+2 c_{G}^{1} \beta_{0}\right) L_{G}+c_G^{2}\right]
\,,\end{aligned}
\ee

where $\Ga_{G}^n=-\frac{j_{G} \kappa_{G}}{2} \Ga_n$. 
$ \Ga_n$, $\gamma_G^n$, $c_G^n$  are coefficients of the universal cusp and non-cusp anomalous dimensions $\Ga_\text{cusp} (\as)$ and $ \gamma_G (\as)$ and of a constant term $c_G$ in $\as$ expansion as
\begin{equation}
    \begin{aligned}
     &\Ga_\text{cusp} (\as) =\sum_{n=0} \Ga_n\,  \left(\frac{\as}{4\pi}\right)^{n+1}
\,,\\
&\gamma_G (\as)=\sum_{n=0} \gamma_G^n\,  \left(\frac{\as}{4\pi}\right)^{n+1}
\label{eq:gin}
\,,\\
&c_G=\sum_{n=1}c_G^n\,  \left(\frac{\as}{4\pi}\right)^{n}
\,,   
    \end{aligned}
\end{equation}
where $\beta_n$ and $\Ga_n$ are given in \appx{anom} and $\gamma_G^n$, and $c_G^n$ are given in \appx{gamma} and \appx{const}, respectively. The factors $j_{G}$ and $\kappa_{G}$ and log $L_G$ are summarized in \tab{jkL}.

\begin{table}[htb]
 $$
\begin{array}{c|cccccc}
\hline
                    & C_t & C_S & \tilde{J} & \tilde{S} \\ \hline
 j_G              & 1    &1        &2-a        &1 \\ \hline
 \kappa_G     & 0    & 2       & -2/(1-a) & 4/(1-a) \\ \hline
L_G   & \ln  \frac{m_t}{\mu_{C_t}} \quad& \ln \frac{i m_H}{\mu_{C_S}} \quad
	& \ln \left[\frac{m_H}{\mu_{\tilde J}}\left(\nu e^{\gamma_{E}}\right)^{-1 / j_{\tilde J}}\right] \quad
	& \ln \left[\frac{m_H}{\mu_{\tilde S}}\left(\nu e^{\gamma_{E}}\right)^{-1 / j_{\tilde S}}\right] \\ \hline
\end{array}
 $$
\caption{ \label{tab:jkL}}
\end{table}

Each term in \eq{str} can be transformed into the momentum space by the inverse Laplace transformation as follows:
\be\label{eq:lap}
\begin{aligned}
1&\rightarrow\delta(\tau)\\-\ln \nu e^{\gamma_E}&\rightarrow\mathcal{L}_0(\tau)\\\ln^2 \nu e^{\gamma_E}&\rightarrow2\mathcal{L}_1(\tau)-\frac{\pi^2}{6}\delta(\tau)\\-\ln^3 \nu e^{\gamma_E}&\rightarrow3\mathcal{L}_2(\tau)-\frac{\pi^2}{2}\mathcal{L}_0(\tau)+2\zeta_3\delta(\tau)\\\ln^4 \nu e^{\gamma_E}&\rightarrow4\mathcal{L}_3(\tau)-2\pi^2\mathcal{L}_1(\tau)+8\zeta_3\mathcal{L}_0(\tau)+\frac{\pi^4}{60}\delta(\tau)
\,,\end{aligned}
\ee
where the distribution function $\cL_n(\tau)$ is defined by
\begin{align}\label{eq:cLn}           
            \cL_n(\tau)&=
            \begin{cases}
            \left[\frac{\ln^n(\tau)}{\tau}\right]_+ & \quad n\ge 0\,, \\
            \delta(\tau) &\quad  n=-1.
 \end{cases}      
\end{align}

\subsection{Non-cusp anomalous dimension}
\label{app:gamma}
The anomalous dimension of the function $G$ is defined by the Renormalization Group (RG) equation as
\be\label{eq:rge}
 \mu \frac{d}{d \mu} G(\nu, \mu)=\gamma_{G}(\mu) G(\nu, \mu)
 \,,\ee
 \be\label{eq:anad}
 \gamma_{G}(\mu)=j_{G} \kappa_{G} \Ga_{\text {cusp }}\left(\alpha_{s}\right) L_{G}+\gamma_{G}\left(\alpha_{s}\right)
 \ee
From scale independence of the factorization in \eq{fac2}, we have a consistence relation:
 \be\label{eq:consre}
\frac{2\beta(\as)}{\as}+2\gamma_{C_t}(\as)+2\gamma_{C_s}(\as)+2\gamma_{J}(\as)+\gamma_{S}(\as)=0
\ee
where the first term is from the Born rate $\Ga_B \propto \as^2$. The non-cusp anomalous dimensions of $C_t$\cite{inami1983effective,djouadi1992higgs,Chetyrkin:1997iv,Chetyrkin:1997un,Chetyrkin:2005ia,baikov2017five} and $C_S$ \cite{Gehrmann:2010ue} are given by
\be\label{eq:ana1}
\begin{aligned}
\gamma_{C_{t}}^{0} &=0 , \quad
\gamma_{C_{t}}^{1} =-2 \beta_{1}\,, \\
\gamma_{C_{S}}^{0} &=0 ,\quad
\gamma_{C_{S}}^{1} =\left(-\frac{118}{9}+4 \zeta_{3}\right) C_{A}^{2}+\left(-\frac{38}{9}+\frac{\pi^{2}}{3}\right) C_{A} \beta_{0}+2 \beta_{1}
\,,\end{aligned}
\ee
The quark soft anomalous dimension is given in \cite{Hornig:2009vb,Bell:2018vaa}, gluon soft anomalous dimension can be obtained from quark soft anomalous dimension by a Casimir scaling that is $C_F\to C_A$. 
\be\label{eq:softana}
\gamma_{S}^{0}(a)=0,\quad
\gamma_{S}^{1}(a)=\frac{2}{1-a}\left[\gamma_{1}^{C A}(a) C_{A}^2+\gamma_{1}^{n f}(a) C_{A} T_{F} n_{f}\right]
\,,\ee
where we have made the $a$-dependence of the anomalous dimension explicit, and
\be\label{eq:coesoftana}
\begin{aligned}
\gamma_{1}^{C A}(a) &=-\frac{808}{27}+\frac{11 \pi^{2}}{9}+28 \zeta_{3}-\Delta \gamma_{1}^{C A}(a) \,,\\
\gamma_{1}^{n f}(a) &=\frac{224}{27}-\frac{4 \pi^{2}}{9}-\Delta \gamma_{1}^{n f}(a)\,,
\end{aligned}
\ee
where

\be\label{eq:intsoftana}
\begin{aligned}
\Delta \gamma_{1}^{C A}(a) &=\int_{0}^{1} d x \int_{0}^{1} d y \frac{32 x^{2}\left(1+x y+y^{2}\right)\left[x\left(1+y^{2}\right)+(x+y)(1+x y)\right]}{y\left(1-x^{2}\right)(x+y)^{2}(1+x y)^{2}} \ln \left[\frac{\left(x^{a}+x y\right)\left(x+x^{a} y\right)}{x^{a}(1+x y)(x+y)}\right], \\
\Delta \gamma_{1}^{n f}(a) &=\int_{0}^{1} d x \int_{0}^{1} d y \frac{64 x^{2}\left(1+y^{2}\right)}{\left(1-x^{2}\right)(x+y)^{2}(1+x y)^{2}} \ln \left[\frac{\left(x^{a}+x y\right)\left(x+x^{a} y\right)}{x^{a}(1+x y)(x+y)}\right],
\end{aligned}
\ee

which vanish for thrust case $a=0$. Here we give numerical result at the values of $a$ used in MC simulation.

\begin{center}
\begin{tabular}{|c||c|c|c|c|c|c|c|}
\hline$a$ & $-1.0$ & $-0.75$ & $-0.5$ & $-0.25$ & $0.0$ & $0.25$ & $0.5$ \\
\hline \hline$\gamma_{1}^{C A}$ & $1.0417$ & $5.8649$ & $9.8976$ & $13.190$ & $15.795$ & $17.761$ & $19.132$ \\
\hline$\gamma_{1}^{n f}$ & $-0.9571$ & $0.5284$ & $1.8440$ & $2.9751$ & $3.9098$ & $4.6398$ & $5.1613$ \\
\hline
\end{tabular}
\end{center}
\begin{multicols}{2}
Then, we can obtain $\gamma_J(\as)$ is given by \eq{consre}.

\end{multicols}
\subsection{Constant term}
\label{app:const}
Content terms in $C_t$ \cite{inami1983effective,djouadi1992higgs,Chetyrkin:1997iv,Chetyrkin:1997un,Chetyrkin:2005ia,baikov2017five} and $C_S$ \cite{Gehrmann:2010ue} are given by
\begin{equation}\label{eq:consC}
\begin{aligned}
&c_{C_t}^1=11,\\
&c_{C_t}^2=\frac{2777}{18}-\frac{67}{6}n_f,\\
&c_{C_S}^1=C_A\frac{\pi^2}{6},\\
&c_{C_S}^2=C_A^2\left(\frac{\pi^{4}}{72}-\frac{143 \zeta_{3}}{9}+\frac{67 \pi^{2}}{36}+\frac{5105}{162}\right)+C_{F} n_{f}\left(8\zeta_3-\frac{67}{6}\right)+C_An_f\left(-\frac{46 \zeta_{3}}{9}-\frac{5 \pi^{2}}{18}-\frac{916}{81}\right)
\,.
\end{aligned}
\end{equation}
The constant term of quark soft functions is known to one-loop for generic values of $a$ analytically \cite{Hornig:2009vb}, the two-loop  constant is determined numerically for generic $a$ in \cite{Bell:2020yzz}. Gluon soft function can be obtained from quark soft function constant by a Casimir scaling that is $C_F\to C_A$. In Laplace space:
\begin{equation}\label{eq:cons}
\begin{aligned}
&c_{\tilde{S}}^{1}(a)=-C_{A} \frac{\pi^{2}}{1-a}\,, \\
&c_{\tilde{S}}^{2}(a)=c_{2}^{C A}(a) C_{A}^2+c_{2}^{n f}(a) C_{A} T_{F} n_{f}+\frac{\pi^{4}}{2(1-a)^{2}} C_{A}^{2}
\,,
\end{aligned}
\end{equation}
where
\begin{center}
\begin{tabular}{|c||c|c|c|c|c|c|c|}
\hline$a$ & $-1.0$ & $-0.75$ & $-0.5$ & $-0.25$ & $0.0$ & $0.25$ & $0.5$ \\
\hline \hline$c_{2}^{C A}$ & $-22.430$ & $-29.170$ & $-36.398$ & $-44.962$ & $-56.499$ & $-74.717$ & $-110.55$ \\
\hline$c_{2}^{n f}$ & $27.315$ & $28.896$ & $31.589$ & $36.016$ & $43.391$ & $56.501$ & $83.670$ \\
\hline
\end{tabular}
\end{center}

One-loop constant of gluon jet function \cite{Ellis:2010rwa,Hornig:2016ahz} in Laplace space is given by

\begin{equation}\label{eq:conj}
\begin{aligned}
\cjone(a)&=c_J^1(a)+\frac{\Gamma_J^0(a)}{(2-a)^2}\frac{\pi^2}6\\
&=\frac{2}{1-a/2}\left[C_A\Bigg((1-a)\Bigg(\frac{67}{18}-\frac{\pi^2}3\Bigg)-\frac{\pi^2}6\frac{(1-a/2)^2-1}{1-a}-f_1(a)\Bigg)-T_F n_f\left(\frac{20-23a}{18}-f_2(a)\right)\right]
\,,
\end{aligned}
\end{equation}
where
\begin{equation}\label{eq:intconj}
\begin{aligned}
    f_1(a)&=\int_0^1dx\frac{\left(1-x(1-x)\right)^2}{x(1-x)}\ln[x^{1-a}+(1-x)^{1-a}]\\
    f_2(a)&=\int_{0}^{1}dx\left(2x(1-x)-1\right)\ln[x^{1-a}+(1-x)^{1-a}]
\end{aligned}
\,.
\end{equation}
Note that in \eq{conj} the minus sign of $\pi^2/6$ term is opposite in Eq.~(11) of Ref.~\cite{Hornig:2016ahz}, which we think is a typo since following their recipe  gives the minus and our numerical result for $\cjone$ is consistent with the minus. 

The two-loop constant is numerically obtained and is given in \tab{c2jgg}.

\section{Resummation formula}
\label{app:res}
The large logs in $C_t$, $C_S$, $\tilde{J}$ and $\tilde{S}$ can be resummed by the RG evolution starting from natural scales, in which the logs are small, to an arbitrary scale $\mu$. The solution of RGE in \eq{rge} shares the following structure:
\begin{equation}\label{eq:ref}
G^\text{res}(\nu, \mu)=G\left(\nu, \mu_{G}\right) e^{K_{G}\left(\mu_{G}, \mu\right)+j_{G} \eta_{G}\left(\mu_{G}, \mu\right) L_{G}}
\,,
\end{equation}
where $K_G$ and $\eta_G$ are integration of $\gamma_G(\mu)$ in \eq{anad} from $\mu_G$ to $\mu$

\be\label{eq:intgam}
\int_{\mu_G}^{\mu}  \frac{d\mu'}{\mu'}\gamma_G(\mu')
= j_G \, \kappa_G \int_{\mu_G}^{\mu} \frac{d\mu'}{\mu'}\Ga_\text{cusp} (\as) \left[-\ln(\mu'/\mu_G)+L_G(\mu_G) \right]+\int_{\mu_G}^{\mu} \frac{d\mu'}{\mu'}\gamma_G (\as)
\,.\ee

After replacing $\tfrac{d \mu'}{\mu'}$ by $d\as/\beta(\as)$, we define three integrals as  
\bea
K_\Ga (\mu_G,\mu)&=&\int^{\as(\mu)}_{\as(\mu_G)} \frac{d\as}{\beta(\as)}\Ga_\cusp (\as )\, \int^{\as}_{\as(\mu_G)} \frac{d\as'}{\beta(\as')}
\,,\nn\\ 
\eta_\Ga (\mu_G,\mu)&=&\int^{\as(\mu)}_{\as(\mu_G)} \frac{d\as}{\beta(\as)}\Ga_\cusp (\as ) 
\,,\nn\\
K_{\gamma_G} (\mu_G,\mu) &=&\int^{\as(\mu)}_{\as(\mu_G)} \frac{d\as}{\beta(\as)}\gamma_G (\as ) 
\,.\label{eq:Keta-def} 
\eea
In \appx{anom} we give results of integration done order by order in $\as$. We define $K_G$ and $\eta_G$ as
\bea
K_G(\mu_G,\mu) &=& -j_G\kappa_G \, K_\Ga (\mu_G,\mu) +K_{\gamma_G} (\mu_G,\mu)
\,,\nn\\
\eta_G (\mu_G,\mu) &=& \kappa_G \, \eta_\Ga (\mu_G,\mu)
\,.\label{eq:ker}
\eea

Now we review conversion of resummed results in Laplace space back to momentum space by the inverse Laplace transformation
\be\label{eq:ilaplace}
\cL^{-1} \left\{ f(\nu) \right\} =  \frac{1}{2\pi i} \int^{\gamma+i\infty}_{\gamma-i\infty} d\nu\, e^{\nu \ta} f(\nu)
\ee
In order to make the inverse integration simple, we replace all $\nu$-dependent log by derivative operators as $L_G\to \partial_{\eta_G}/j_G$ in the fixed-order function $G(\nu,\mu)$ in \eq{str} and rewrite the resummed function in \eq{ref} as
\be\label{eq:GLG}
G^\text{res}(\nu,\mu)= g(\partial_{\eta_G})\, e^{K_G(\mu_G,\mu)+j_G \eta_G(\mu_G,\mu) L_G} 
\,,\ee
 where $g(\partial_{\eta_G})=G(\nu,\mu)\vert_{L_G\to \partial_{\eta_G}/j_G}$. One finds \eq{GLG} is equivalent to \eq{ref}.

At $\mathcal{O}(\alpha_s)$ we have the fix-order function $g(\partial_{\eta_G})$:
\be\label{eq:gas}
g(\partial_{\eta_G})=1+\frac{\as(\mu)}{4\pi}\left[ -\kappa_G \frac{\Ga_0}{2j_G}\partial^2_{\eta_G} -\frac{\gamma^G_0}{j_G} \partial_{\eta_G} +c^1_G \right]
\,,
\ee
Using an identity of inverse transformation 
\be\label{eq:Linv}
\cL^{-1} \left\{ \nu^{-\eta_G}  \right\}= \frac{\tau_a^{\eta_G-1}}{\Ga(\eta_G)} 
\,,\ee
\eq{GLG} turns into the resummed function in momentum space as
\be\label{eq:Gtaua}
G^\text{res}(\tau_a,\mu)
=\frac{e^{K_G}}{\tau_a}\,  g(\partial_{\eta_G})\frac{e^{j_G \eta_G L_G(\tau_a)}}{ \Ga(\eta_G)}
\,,\ee
where the logarithm in $\ta$ is defined by
\be
L_G(\tau_a)=\ln\left[ \frac{Q}{\mu_G} \left(\ta e^{-\gamma_E}\right)^{1/j_G} \right] 
\,,
\qquad G=\{S,J\}
\,.
\label{eq:LGtau}\ee

The resummed distribution is produc of jet, and soft functions expressed in a form of \eq{GLG} then, $\eta_G$ in \eq{Linv} is replaced by the sum $\Omega=\eta_S+2\eta_J$. Finally, we obtain:

\begin{equation}\label{eq:drnu}
\begin{aligned}
\frac{1}{\Ga_B}\frac{d \Ga^\text{res}}{d \ta}=
& \frac{\as(\mu)^2}{\as(m_H)^2}\left|C_t\left(m_t,\mu_{C_t}\right)\right|^2\left|C_s\left(m_H,\mu_H\right)\right|^2 j^2\left(\partial_{\eta_{J}}\right) s\left(\partial_{\eta_{S}}\right) \\
& \times e^{\kappa\left(\left\{\mu_{i}\right\}, \mu\right)}\left(\frac{m_t}{\mu_{C_t}}\right)^{2\eta_{C_t}\left(\mu_{C_t}, \mu\right)}\left|\bigg( \frac{i m_H}{\mu_{C_s}}\bigg)^{\eta_{C_s}(\mu_{C_s},\mu)} \right|^2\left(\frac{m_H}{\mu_{J}}\right)^{2j_{J} \eta_{J}\left(\mu_{J}, \mu\right)}\left(\frac{m_H}{\mu_{S}}\right)^{j_{S} \eta_{S}\left(\mu_{S}, \mu\right)} \\
& \times \frac{\tau_{a}^{-1+\Omega}}{\Ga(\Omega)} e^{-\gamma_{E} \Omega} 
\,,
\end{aligned}
\end{equation}
where 
\be
\kappa(\{\mu_i\},\mu)=2K_{C_t}(\mu_{C_t},\mu)+K_{C_S}(\mu_{C_S},\mu)+K_{C_S}^*(\mu_{C_S},\mu)+2K_J(\mu_J,\mu)+K_S(\mu_S,\mu)
\,.
\ee
In the numerical calculation, we set $\mu_{C_t}=\mu_{C_S}$ to be the hard scale $\mu_H$ and vary them at the same time for a scale variation. 
By shifting the derivative by $L_G(\ta)$ all the exponents in \eq{Gtaua} can be moved in the front as 
\be\label{eq:Gtaua2}
G^\text{res}(\ta,\mu)
=\frac{e^{K_G+j_G \eta_G L_G(\ta)}}{\ta}\,  g\left(\partial_{\Omega}+j_G L_G(\tau_a) \right)\frac{1}{\Ga(\Omega)}
\,.\ee

The operators $[\partial_{\Omega}+j_G L_G(\ta)]^n $ turns into poly-logarithms
\be\label{eq:invop}
\begin{aligned}
\left[\partial_{\Omega}+j_GL_G\left(\tau_{a}\right)\right] \frac{1}{\Ga\left(\Omega\right)} &=\psi_G\left(\Omega\right) \frac{1}{\Ga\left(\Omega\right)} 
\,,\\
\left[\partial_{\Omega}+j_GL_G\left(\tau_{a}\right)\right]^{2} \frac{1}{\Ga\left(\Omega\right)} &=\left\{\psi_G^{2}\left(\Omega\right)-\psi^{(1)}\left(\Omega\right)\right\} \frac{1}{\Ga\left(\Omega\right)} 
\,,\\
\left[\partial_{\Omega}+j_GL_G\left(\tau_{a}\right)\right]^{3} \frac{1}{\Ga\left(\Omega\right)} &=\left\{\psi_G^{3}\left(\Omega\right)-3 \psi_G \psi^{(1)}\left(\Omega\right)-\psi^{(2)}\left(\Omega\right)\right\} \frac{1}{\Ga\left(\Omega\right)} 
\,,\\
\left[\partial_{\Omega}+j_GL_G\left(\tau_{a}\right)\right]^{4} \frac{1}{\Ga\left(\Omega\right)} &=\left\{\psi_G^{4}\left(\Omega\right)-6 \psi_G^{2} \psi^{(1)}\left(\Omega\right)-4 \psi_G \psi^{(2)}\left(\Omega\right)+3\left(\psi^{(1)}\left(\Omega\right)\right)^{2}-\psi^{(3)}\left(\Omega\right)\right\} \frac{1}{\Ga\left(\Omega\right)}
\,,\end{aligned}
\ee
where 
\be\label{eq:invopw}
\begin{aligned}
 &\psi(\Omega)=\Ga'(\Omega)/\Ga(\Omega)
 \,,\\
 &\psi_G\left(\Omega\right)=-\psi(\Omega)+j_GL_G(\tau_a)
\,,\\ 
&\psi^{(n)}(\Omega)=\frac{d^n\psi(\Omega)}{d\Omega^n}
\,.\end{aligned}
\ee
We would like to note that operators in crossing terms like $\as^2 s_1j_1$ are not simply the square of the second line in \eq{invop} and need to be carefully expended as
\be\label{eq:invop11}
\begin{aligned}
&\left[\partial_{\Omega}+j_{G1} L_{G1}\left(\tau_{a}\right)\right]^2 \left[\partial_{\Omega}+j_{G2} L_{G2}\left(\tau_{a}\right)\right]^2\frac{1}{\Ga\left(\Omega\right)}
\\
&=\left\{\psi_{G1}^2\psi_{G2}^2-\left(\psi_{G1}^2+4\psi_{G1}\psi_{G2}+\psi_{G2}^2\right)\psi^{(1)}(\Omega)\right.
\left.-\left(2\psi_{G1}+2\psi_{G2}\right)\psi^{(2)}\left(\Omega\right)+3\left(\psi^{(1)}\left(\Omega\right)\right)^2-\psi^{(3)}\left(\Omega\right)\right\}\frac{1}{\Ga(\Omega)}
\end{aligned}
\ee


Our final form of the resummed result is given by
\bea 
\frac{1}{\Ga_0}\frac{d \Ga^\text{res}}{d \tau_{a}}
&=& 
\frac{\as(\mu)^2}{\as(m_H)^2}\left|C_t\left(m_t,\mu_{C_t}\right)\right|^2\left|C_s\left(m_H,\mu_{C_S}\right)\right|^2 ~
e^{\kappa(\{\mu_i\},\mu)}
\nn\\
& &\times 
\ta^{-1+\Omega}e^{-\gamma_E \Omega} ~ 
\left(\frac{m_t}{\mu_{C_t}}\right)^{2\eta_{C_t}\left(\mu_{C_t}, \mu\right)}
\left|\bigg( \frac{i m_H}{\mu_{C_S}}\bigg)^{\eta_{C_S}(\mu_{C_S},\mu)} \right|^2
\bigg(\frac{m_H}{\mu_J}\bigg)^{2j_J\eta_J(\mu_J,\mu)} 
\bigg(\frac{m_H}{\mu_S}\bigg)^{j_S\eta_S(\mu_S, \mu)}
\nn\\
&& \times
\bigg[
j^2\bigg(\partial_{\Omega}+j_J L_J(\ta),\mu_J\bigg)
s\bigg(\partial_{\Omega} +j_S L_S(\ta),\mu_S\bigg)  
 \bigg]
\frac{1}{ \Ga(\Omega)}
 \,. 
\label{eq:sigta2}
\eea

\section{Cusp anomalous dimensions and related integral} 
\label{app:anom}

Here, we summarize the coefficient of cusp-anomalous and beta function and integrals defined in \eq{Keta-def} in the resummation.

To the NNLL order, we need up to three-loop cusp anomalous dimension for gluon~\cite{Moch:2004pa,Korchemsky:1987wg}, which is related to quark channel by the ratio of the eigenvalues $C_i$ of the quadratic Casimir operators up to 3-loop \cite{Moch:2004pa}: $\frac{\Gamma_{\mathrm{cusp}}^q(\alpha_s)}{C_F}=\frac{\Gamma_{\mathrm{cusp}}^g(\alpha_s)}{C_A}$. Note that it is different from the soft function where all the $C_F$ are replaced by $C_A$. Here we give cusp anomalous dimension for gluon without subscript $g$.

\begin{subequations}
\begin{align}
\Ga_0 &= 4 C_A \nn\\
\Ga_1 &= \Ga_0 \Bigl[ \Bigl( \frac{67}{9} -\frac{\pi^2}{3}\Bigr) C_A - \frac{20}{9} T_F n_f\Bigr] \nn\\
\Ga_2 &= \Ga_0 \Bigl[ \Bigl( \frac{245}{6} - \frac{134\pi^2}{27} + \frac{11\pi^4}{45} + \frac{22\zeta_3}{3} \Bigr) C_A^2 + \Bigl( - \frac{418}{27} + \frac{40\pi^2}{27} - \frac{56\zeta_3}{3}\Bigr) C_A T_F n_f  + \Bigl(-\frac{55}{3} + 16\zeta_3\Bigr) C_F T_F n_f - \frac{16}{27} T_F^2 n_f^2\Bigr]
\end{align}
\end{subequations}

The beta function expanded in powers of $\as$ is given by
\be
\beta(\alpha_s) =\mu \frac{d \as(\mu)}{d\mu}
=- 2 \alpha_s \sum_{n=0}^\infty \beta_n\Bigl(\frac{\alpha_s}{4\pi}\Bigr)^{n+1}
\ee
The beta-function coefficients~\cite{larin1993three,tarasov1980gell} are 
\begin{align} \label{eq:betai}
\beta_0 &= \frac{11}{3}\,C_A -\frac{4}{3}\,T_F\,n_f
\,,\nn\\
\beta_1 &= \frac{34}{3}\,C_A^2  - \Bigl(\frac{20}{3}\,C_A\, + 4 C_F\Bigr)\, T_F\,n_f
\,, \nn\\
\beta_2 &=
\frac{2857}{54}\,C_A^3 + \Bigl(C_F^2 - \frac{205}{18}\,C_F C_A
 - \frac{1415}{54}\,C_A^2 \Bigr)\, 2T_F\,n_f
 + \Bigl(\frac{11}{9}\, C_F + \frac{79}{54}\, C_A \Bigr)\, 4T_F^2\,n_f^2
\,.\end{align}

An expansion of \eq{Keta-def} in powers of $\as$ is
\begin{align} 
K_\Ga(\mu_0, \mu) &= -\frac{\Ga_0}{4\beta_0^2}\,
\biggl\{ \frac{4\pi}{\alpha_s(\mu_0)}\, \Bigl(1 - \frac{1}{r} - \ln r\Bigr)
   + \biggl(\frac{\Ga_1 }{\Ga_0 } - \frac{\beta_1}{\beta_0}\biggr) (1-r+\ln r)
   + \frac{\beta_1}{2\beta_0} \ln^2 r
\nn\\ & \quad
+ \frac{\alpha_s(\mu_0)}{4\pi}\, \biggl[
  \biggl(\frac{\beta_1^2}{\beta_0^2} - \frac{\beta_2}{\beta_0} \biggr) \Bigl(\frac{1 - r^2}{2} + \ln r\Bigr)
  + \biggl(\frac{\beta_1\Ga_1 }{\beta_0 \Ga_0 } - \frac{\beta_1^2}{\beta_0^2} \biggr) (1- r+ r\ln r)
  - \biggl(\frac{\Ga_2 }{\Ga_0} - \frac{\beta_1\Ga_1}{\beta_0\Ga_0} \biggr) \frac{(1- r)^2}{2}
     \biggr] \biggr\}
\,, \nn\\
\eta_\Ga(\mu_0, \mu) &=
 - \frac{\Ga_0}{2\beta_0}\, \biggl[ \ln r
 + \frac{\alpha_s(\mu_0)}{4\pi}\, \biggl(\frac{\Ga_1 }{\Ga_0 }
 - \frac{\beta_1}{\beta_0}\biggr)(r-1)
 + \frac{\alpha_s^2(\mu_0)}{16\pi^2} \biggl(
    \frac{\Ga_2 }{\Ga_0 } - \frac{\beta_1\Ga_1 }{\beta_0 \Ga_0 }
      + \frac{\beta_1^2}{\beta_0^2} -\frac{\beta_2}{\beta_0} \biggr) \frac{r^2-1}{2}
    \biggr]
\,, \nn\\
K_\gamma(\mu_0, \mu) &=
 - \frac{\gamma_0}{2\beta_0}\, \biggl[ \ln r
 + \frac{\alpha_s(\mu_0)}{4\pi}\, \biggl(\frac{\gamma_1 }{\gamma_0 }
 - \frac{\beta_1}{\beta_0}\biggr)(r-1) \biggr]
\label{eq:Keta}
\,.\end{align}
Here, $r = \alpha_s(\mu)/\alpha_s(\mu_0)$. Solving the beta function to three-loop order gives the running coupling expressed by
\begin{align} \label{eq:alphas}
\frac{1}{\alpha_s(\mu)} &= \frac{X}{\alpha_s(\mu_0)}
  +\frac{\beta_1}{4\pi\beta_0}  \ln X
  + \frac{\alpha_s(\mu_0)}{16\pi^2} \biggr[
  \frac{\beta_2}{\beta_0} \Bigl(1-\frac{1}{X}\Bigr)
  + \frac{\beta_1^2}{\beta_0^2} \Bigl( \frac{\ln X}{X} +\frac{1}{X} -1\Bigr) \biggl]
\,,\end{align}
where $X\equiv 1+\alpha_s(\mu_0)\beta_0 \ln(\mu/\mu_0)/(2\pi)$.
In our numerical calculations, we take the full NNLL results in \eq{Keta} for $K_{\Ga,\gamma},\eta_\Ga$ and in \eq{alphas}. 
\begin{multicols}{2}

\section{Tikhonov method} 
\label{app:Tikhonov}

For a known matrix $A$ and vector $\mathbf{b}$, a least-square method is to find a vector $\mathbf{x}$ such that it minimizes the square residuals $|A\mathbf{x}-\mathbf{b} |^2$, where $|\cdot|^2$ is the vector squared. 
In our problem, $\mathbf{b}_i$ is the value of remainder function at $i_\text{th}$ bin, $\mathbf{x}_j$ is $j_\text{th}$ parameter in the fit function, and $A_{ij}$ is the value of $j_\text{th}$ term of fit function at $i_\text{th}$ bin.
$\hat{x}=(A^T A)^{-1}A^T\mathbf{b}$ is the solution to zero gradient of the square residuals.
However, in some cases, matrix $A^T A$ is nearly singular and sensitive to small changes in data like noise. In those cases regression is said to be ill-posed, and generally regularization helps improve the result. We adopt Tikhonov regularization \cite{tikhonov1963solution}, in which a regularization term is included in the squared residuals $|A\mathbf{x}-\mathbf{b}|^2+\lambda | \Gamma\mathbf{x}|^2$, where $\lambda$ is a regularization parameter and $\Gamma$ is a Tikhonov matrix. With the introduction of the regularization term, the solution to zero gradient is $\hat{x}=(A^T A+\lambda\Gamma^T\Gamma)^{-1}A^T\mathbf{b}$, where $\lambda\Gamma^T\Gamma$ tames the singularity of $A^T A$. 

The regularization depends on the choice of regularization parameter $\lambda$ and Tikhonov matrix $\Gamma$. We take the identity matrix for $\Gamma$ in our analysis. We also test with discrete first- and second-order derivative matrices that give $\Gamma\mathbf{x}=(x_2-x_1,\cdots, x_n-x_{n-1})^T$ and $(x_3-2x_2+x_1,\cdots, x_n+2x_{n-1}-x_{n-2})^T$ and they essentially give consistent result with that with identity matrix. The regularization parameter should be assigned carefully, a tiny value cannot tame the singularity, while a too large value would distort the feature of the remainder function. 
We select the value of parameter that gives a robust fit result against variations of upper and lower bounds of the fit region. 
We find that for non-positive $a$ cases, the value of the regularization parameter is small $\lambda=10^{-4}$, which essentially equivalent to ordinary least-square, i.e., $\lambda= 0$. For positive $a$, the parameter is larger $\lambda= 10^3(10^1)$ for $a=0.25$ and $\lambda= 10^4(10^3)$ for $a=0.5$ in the quark (gluon) channel.

\end{multicols}
\begin{figure}[!htbp]
\centering
\includegraphics[scale=0.35]{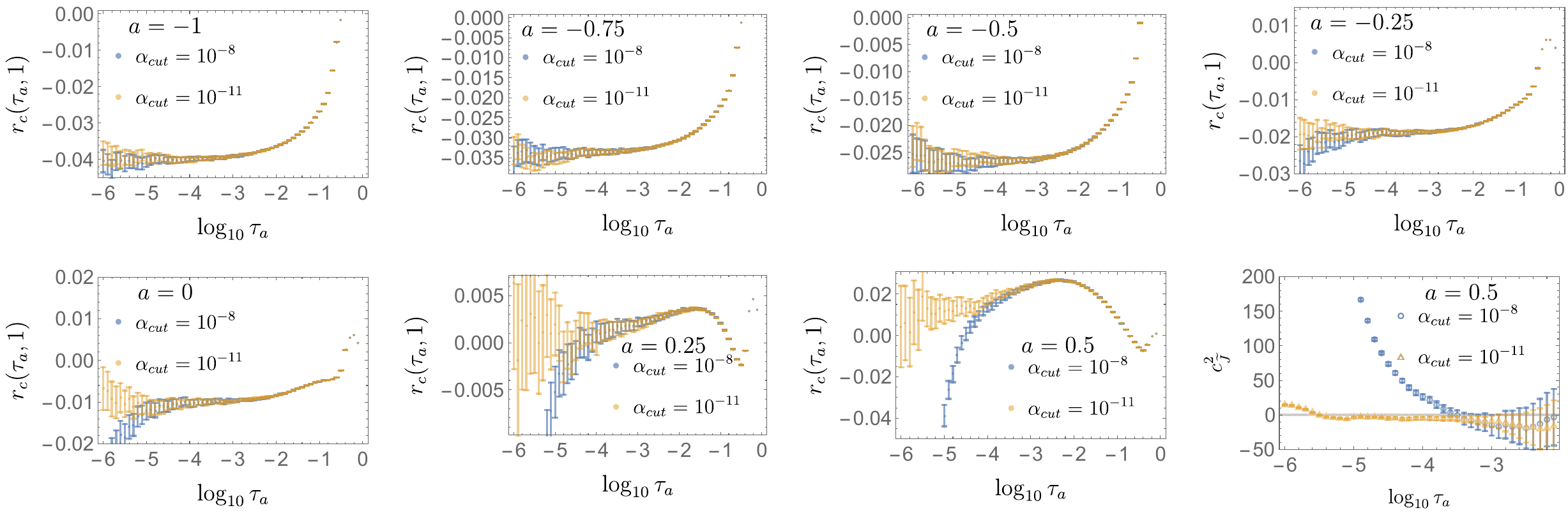}
\caption{NLO remainder functions for $\alpha_\text{cut}=10^{-11}$ (blue) and for $10^{-8}$ (yellow) at seven different values of $a$. The last plot shows prediction of 2-loop constant $\cjt$ as a function of lower bound $\ta^\text{low}$ for $\alpha_\text{cut}=10^{-11}$ and for $10^{-8}$.} 
\label{fig:yijk}
\end{figure}
\begin{multicols}{2}

\section{Cutoff effect} 
\label{app:cutoff}

We explore the effect of cutoff $\alpha_\text{cut}$ in the quark channel by comparing our standard value $\alpha_\text{cut}=10^{-11}$ and larger value $10^{-8}$ as shown in \fig{yijk}. 
Their differences are hardly visible in $a< 0$ in given ranges of $\ta$ and become visible in $a\ge0$. 
The last plot in \fig{yijk} shows the constant $\cjt$ as a function of $\ta^\text{low}$ at $a=0.5$. While the value of constant stays still in the domain $\log_{10} \ta^\text{low}\in\{-4.5,-2.6\}$ for $\alpha_\text{cut}=10^{-11}$,  for $\alpha_\text{cut}=10^{-8}$ it begins to change around $\log_{10}\ta^\text{low}=-3.5$  and below. This implies that the fit regions for $\alpha_\text{cut}=10^{-11}$ in \fig{hqq_nlo} is insensitive to the cutoff effect.
For $\alpha_\text{cut}=10^{-8}$ we use same criteria and obtained $\cjt=-55.90\pm1.84$ at $a=0.25$ and $-10.33\pm11.44$ at $a=0.5$, which are consistent with the values in \tab{c2jqq}.

\begin{figure}[H]
\centering
\includegraphics[scale=0.35]{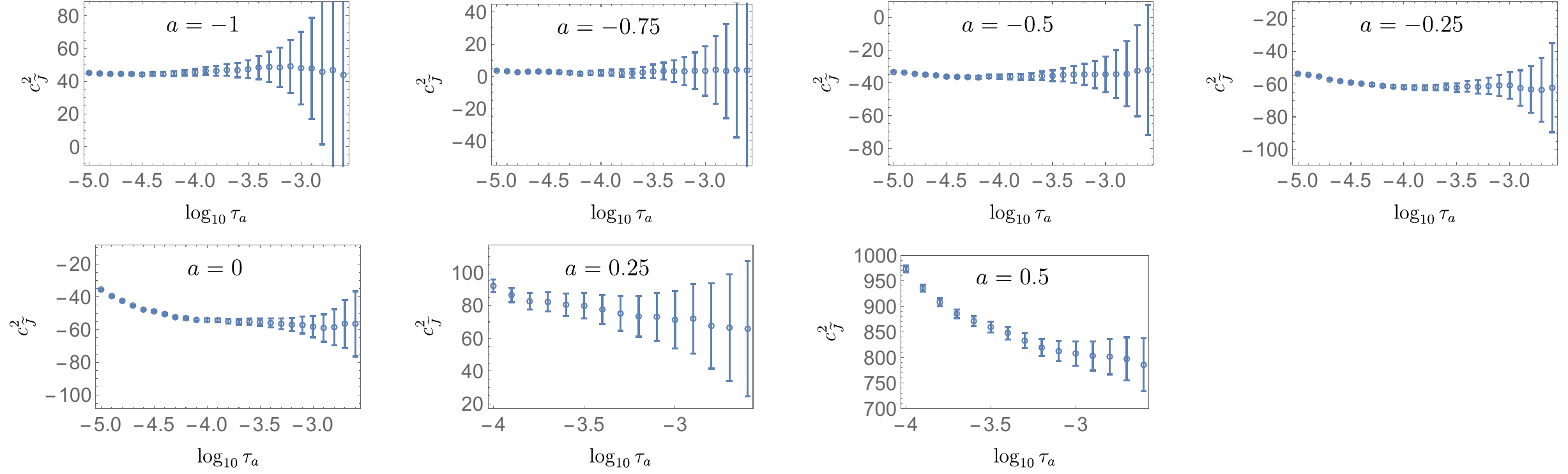}
\caption{
two-loop constant $\cjt$ as a function of lower bound $ \ta^\text{low}$ with fixed-upper bound $\log_{10} \ta^\text{high}=-1.4+0.4\,(1+a)$
at $\alpha_\text{cut}=10^{-8}$
}
\label{fig:hgg_lb}
\end{figure}

\fig{hgg_lb} shows the cutoff effect in the gluon channel as a function of $\ta^\text{low}$ with $\log_{10}\ta^\text{high}=-1.4+0.4\,(1+a)$ held fixed.
Similar to the quark channel, larger $a$ values tend to suffer from greater cutoff effect  $\cjt$. Using the results, our fit regions are selected to avoid the  cutoff effect as described in the main body of the paper.

\end{multicols}
\begin{multicols}{2}

{ }

\end{multicols}

\clearpage

\end{document}